\documentclass{jfm}
\usepackage{natbib,amssymb,amsmath,upmath}
\usepackage{graphicx}
\usepackage{color}
\usepackage[sfdefault=cmbr]{isomath}
\def\v{\vectorsym}
\def\t{\tensorsym}

\providecommand\bnabla{\boldsymbol{\nabla}}
\providecommand\bcdot{\boldsymbol{\cdot}}

\newsavebox{\astrutbox}
\sbox{\astrutbox}{\rule[-5pt]{0pt}{20pt}}

\def\v{\boldsymbol}

\include{help}
\title[Transport of a dilute active suspension]{Transport of a dilute active suspension in pressure-driven channel flow}

\author[B.\ Ezhilan and D.\ Saintillan]%
{Barath Ezhilan and David Saintillan\thanks{Email address for correspondence: dstn@ucsd.edu}}

\affiliation{Department of Mechanical and Aerospace Engineering, University of California San Diego, 9500 Gilman Drive, La Jolla, CA 92093-0411, USA}

\date{\today}

\pubyear{2013}
\volume{??}
\pagerange{??--??}
\date{\today}
\usepackage{graphicx,color,bm}
\graphicspath{{converted_graphics/}}
\usepackage{graphicx}
\begin{document}

\maketitle

\begin{abstract}
Confined suspensions of active particles show peculiar dynamics characterized by wall accumulation, as well as upstream swimming, centerline depletion and shear-trapping when a pressure-driven flow is imposed. We use theory and numerical simulations to investigate the effects of confinement and non-uniform shear on the dynamics of a dilute suspension of \textcolor{black}{Brownian} active swimmers by incorporating a detailed treatment of boundary conditions within a simple kinetic model where the configuration of the suspension is described using a conservation equation for the probability distribution function of  particle positions and orientations, \textcolor{black}{and where particle-particle and particle-wall hydrodynamic interactions are neglected.}  Based on this model, we first investigate the effects of confinement in the absence of flow, in which case the dynamics is governed by a swimming P\'eclet number, or ratio of the persistence length of particle trajectories over the channel width, and \textcolor{black}{a second swimmer-specific parameter whose inverse measures the strength of propulsion. In the limit of weak and strong propulsion, asymptotic expressions for the full distribution function are derived. For finite propulsion, analytical expressions for the concentration and polarization profiles are also obtained using a truncated moment expansion of the distribution function.} In agreement with experimental observations, the existence of a  concentration/polarization boundary layer in wide channels is reported and characterized, suggesting that wall accumulation in active suspensions is primarily a kinematic effect which does not require hydrodynamic interactions. Next, we show that application of a pressure-driven Poiseuille flow leads to net upstream swimming of the particles relative to the flow, and an analytical expression for the mean upstream velocity is derived in the weak flow limit. In stronger imposed flows, we also predict the formation of a depletion layer near the channel centerline, due to cross-streamline migration of the swimming particles towards high-shear regions where they become trapped, and an asymptotic analysis in the strong flow limit is used to obtain a scale for the depletion layer thickness and to rationalize the non-monotonic dependence of the intensity of depletion upon flow rate. Our theoretical predictions are all shown to be in excellent agreement with finite-volume numerical simulations of the kinetic model, and are also supported by recent experiments on bacterial suspensions in microfluidic devices. 

\end{abstract}


\section{Introduction}

The interaction of active self-propelled particles with rigid boundaries under confinement plays a central role in many biological processes. Spermatozoa are well known to accumulate at rigid boundaries \citep{Rothschild63,Woolley03}, with complex implications for their transport in the female tract during mammalian reproduction \citep{Suarez06,Denissenko12,Kantsler14}. The aggregation of bacteria near surfaces and their interaction with external flows in confinement has a strong effect on their ability to adhere and form biofilms \citep{Rusconi10,Lecuyer11,Kim14}. It also impacts their interactions with the gastrointestinal wall during digestion, with consequences for various pathologies \citep{Lu01,Cellia09}. Confinement has also been shown to affect cell-cell interactions and collective motion in dense sperm and bacterial suspensions  and can also result in spontaneous unidirectional flows \citep{Riedel05,Wioland13,Lushi14}. In engineering, the ability to concentrate or separate bacteria by controlling their motions in microfluidic devices with complex geometries has been demonstrated \citep{Galajda07,Hulme08,Lambert10,Kaiser12,Altshuler13}, as well as the ability to harness bacterial swimming power to actuate gears \citep{Sokolov10,Dileonardo10} or transport cargo \citep{Koumakis13,Kaiser14}. Particle-wall interactions are also critical in systems involving synthetic microswimmers \citep{Gibbs11,Takagi13,Takagi14}, as these inherently reside near surfaces due to sedimentation. \vspace{0.02cm}

The prominent feature of confined active suspensions is the tendency of swimming particles to accumulate near  boundaries. This was first brought to light by \cite{Rothschild63}, who measured the concentration of swimming bull spermatozoa in a glass chamber and reported a nonuniform distribution across the channel with a strong spike in concentration near the walls. \cite{Berke08} repeated the same experiment using suspensions of \textit{Escherichia coli}  in microchannels and also observed an  accumulation of bacteria at the channel walls. They further reported  the tendency of bacteria to align parallel to the boundaries, which led them to consider wall hydrodynamic interactions due to the force dipole exerted on the fluid by the self-propelled particles as a potential mechanism for migration. Hydrodynamic interactions are indeed known to have an impact on the trajectories of swimming particles near no-slip walls \citep{Lauga06,Spagnolie12}, and have been shown to lead to attraction of sperm cells towards walls \citep{Fauci95}. \cite{Li09} and \cite{Li11} also observed wall accumulation in suspensions of \textit{Caulobactor crescentus} but presented an alternate mechanism based purely on kinematics that explains accumulation as a result of the collisions of the bacteria with the wall, leading to their reorientation parallel to the surface. The possibility of a non-hydrodynamic mechanism for wall accumulation is indeed supported by various simulations that neglected wall hydrodynamic interactions \citep{Costanzo12,Elgeti13}, suggesting that such interactions in fact only play a secondary role in this process. \vspace{0.02cm}

Several other interesting effects have also been reported when an external flow is applied on the suspension. One such effect is the propensity of motile particles to swim upstream in a pressure-driven flow. This was noted for instance by \cite{Hill07}, who tracked the trajectories of \textit{Escherichia coli} in a shear flow near a rigid surface in a microfluidic channel, and proposed a complex mechanism for upstream swimming based on the chirality of the flagellar bundles and on hydrodynamic interactions. Such interactions were characterized more precisely by \cite{Kaya09}, who demonstrated that the \textit{Escherichia coli} cells undergo modified Jeffery's orbits \citep{Jeffery22} near the walls and suggested that this detail is crucial in understanding the upstream migration. A clearer picture of this phenomenon emerged in yet more recent work by \cite{Kaya12}, who systematically analyzed \textit{Escherichia coli} motility near a surface as a function of the local shear rate. At low shear rates, circular trajectories were observed due to the chirality of the cells, as previously explained by \cite{Lauga06}. At higher shear rates, positive rheotaxis was reported and accompanied by rapid and continuous upstream motility. This directional swimming was explained as a result of the combined effects of surface hydrodynamic interactions, which were thought to cause the swimming cells to dip towards the walls, and of reorientation by the shear flow, which aligns the cells against the flow. Upstream motility was also recently discussed by \cite{Kantsler14} in the case of mammalian spermatozoa, where the combination of shear alignment, wall steric interactions and cell chirality was shown to lead to steady spiraling trajectories in cylindrical capillaries. \vspace{0.1cm}

\vspace{0.0cm}While most experimental studies under confinement have focused on near-wall aggregation and swimming dynamics, the behavior of self-propelled micro-organisms under flow in the bulk of the channels is also of interest. In recent work, \cite{Rusconi14} analyzed the effects of a Poiseuille flow on the trajectories and distributions of motile \textit{Bacilus subtilis} cells, with focus on the central portion of the channel. In sufficiently strong flow, they reported the formation of a depletion layer in the central low-shear region of the channel, accompanied by cell trapping in the high-shear regions surrounding the depletion. This trapping was attributed to the strong alignment of the swimming cells with the flow under high shear, which hinders their ability to swim across streamlines. Quite curiously, they reported that maximum depletion is achieved at a critical imposed shear rate of approximately 10 s$^{-1}$, above which both trapping and depletion become weaker. A simple Langevin model capturing the effects of self-propulsion, shear rotation, and diffusion was also proposed to explain these observations, and was able to reproduce the salient features of the experiments.\vspace{0.1cm}


\vspace{0.0cm}Models and simulations explaining the mechanisms leading to these rich dynamics have been relatively scarce. Direct numerical simulations of hydrodynamically interacting swimming particles confined to a gap between two plates were first performed by \cite{Hernandez05,Hernandez09} using a simple dumbbell model, and indeed captured a strong particle accumulation at the  boundaries in dilute systems. As the mean swimmer density was increased, collective motion and mixing due to particle-particle hydrodynamic interactions led to a decrease in the concentration near the walls. Accumulation was also observed in simulations of self-propelled spheres by \cite{Elgeti13}, who entirely neglected hydrodynamic interactions. This study, as mentioned above, suggests that wall hydrodynamic interactions are not required to explain migration, and neither is shape anisotropy. Rather, the simple combination of cell swimming, steric exclusion by the walls, and diffusive processes is sufficient to capture accumulation, and \cite{Elgeti13} also proposed a simple Fokker-Planck description of the suspension that shares similarities with the present work and was able to explain their results. A similar continuum model was also proposed by \cite{Lee2013}, who derived analytical expressions for the ratio of particles in the bulk vs near-wall region in the limits of weak and strong rotational diffusion. Very recently, \cite{Li14} performed direct numerical simulations of confined  suspensions of spherical squirmers that propel via an imposed slip velocity, and reported strong accumulation at the boundaries irrespective of the details of propulsion. They also noted the tendency of particles  to align normal to the wall in the near-wall region. \vspace{0.1cm}

\vspace{0.0cm}The effects of an external flow have also been addressed using discrete particle models and simulations. The dynamics of isolated deterministic microswimmers in Poiseuille flow were studied in detail by \cite{Zottl12,Zottl13}, who found that such swimmers perform either an upstream-oriented periodic swinging motion or a periodic tumbling motion depending on their location in the channel. Suspensions of interacting swimmers in pressure-driven flow have also been simulated, notably by \cite{Nash10} and \cite{Costanzo12}, who both observed aggregation at the walls together with upstream swimming as a result of the rotation of the particles by the flow. More recently, \cite{Chilukuri14} extended the simulation method of \cite{Hernandez09} to account for a Poiseuille flow. Similar trends as reported earlier were observed, including wall accumulation and upstream swimming, as well as the reduction of accumulation with increasing flow rate. In addition, they also reported the formation of a depletion layer near the channel centerline in strong flows, in agreement with the microfluidic experiments of \cite{Rusconi14}. Simple scalings for the dependence of this depletion with shear rate, swimming speed and channel width were also proposed.

While these various numerical simulations have been able to reproduce the relevant features of previous experiments, a clear unified theoretical model capable of capturing and explaining all of the above effects based on conservation laws and microscopic swimmer dynamics is still lacking. In unconfined systems, much progress has been made over the last decade in the description of the behavior of active suspensions using continuum kinetic theories \citep{SS2013,Marchetti13,Subra09}. One such class of models, introduced by \cite{SS2008a,SS2008b} to explain the emergence of collective motion in semi-dilute suspensions, is based on a conservation equation for the distribution function $\Psi(\boldsymbol{x},\boldsymbol{p},t)$ of particle positions and orientations, in which fluxes arise due to self-propulsion, advection and rotation by the background fluid flow, as well as diffusive processes. When coupled to a model for the fluid flow (whether externally imposed or driven by the swimmers themselves), this conservation equation can be linearized for the purpose of a stability analysis or integrated in time to investigate nonlinear dynamics. This approach, which also relates to other models developed in the context of active liquid crystals \citep{Baskaran09,Marchetti13,Forest13}, has been very successful at elucidating the mechanisms leading to collective motion at a suspension level. \textcolor{black}{However, attempts to apply such continuum kinetic theories to confined suspensions have been few and far between, in part due to the complexity of the boundary conditions that need to be enforced on the distribution function.} 

In this paper, we present a simple continuum theory for the dynamics and transport of a dilute suspension of \textcolor{black}{Brownian} active swimmers in a pressure-driven channel flow between two parallel flat plates. To focus on the effects of steric confinement and its interaction with the flow, we neglect particle-particle and particle-wall hydrodynamic interactions entirely but incorporate a detailed treatment of the boundary conditions for the distribution function. As we show below, our theory is able to capture all the different regimes discussed above, including wall accumulation in the absence of flow, and upstream swimming, depletion at the centerline and trapping in high-shear regions when a flow is applied. We introduce the governing equations, boundary conditions and nondimensionalization in \S \ref{sec:goveq}, where we also derive a simpler approximate model based on moment equations. The equilibrium distributions in the absence of flow are obtained in \S \ref{sec:noflow}, where wall accumulation is seen to be accompanied by a net polarization of the particle distribution near the boundaries, and where a very simple expression is derived for the concentration profile across the channel in terms of the parameters of the problem. The effects of an external Poiseuille flow are discussed in \S \ref{sec:flow}, where a numerical solution of the governing equations captures upstream swimming and shear trapping in the relevant parameter ranges, and where both effects are also explained theoretically using asymptotic analyses in the weak and strong flow regimes. We summarize our results in \S \ref{sec:conclusions} and discuss them in the light of the recent literature in the field.

\section{Governing equations\label{sec:goveq}}

\subsection{Problem definition and kinetic model\label{sec:model}}

\begin{figure}
\centering\vspace{0.3cm}
\includegraphics[width=12cm]{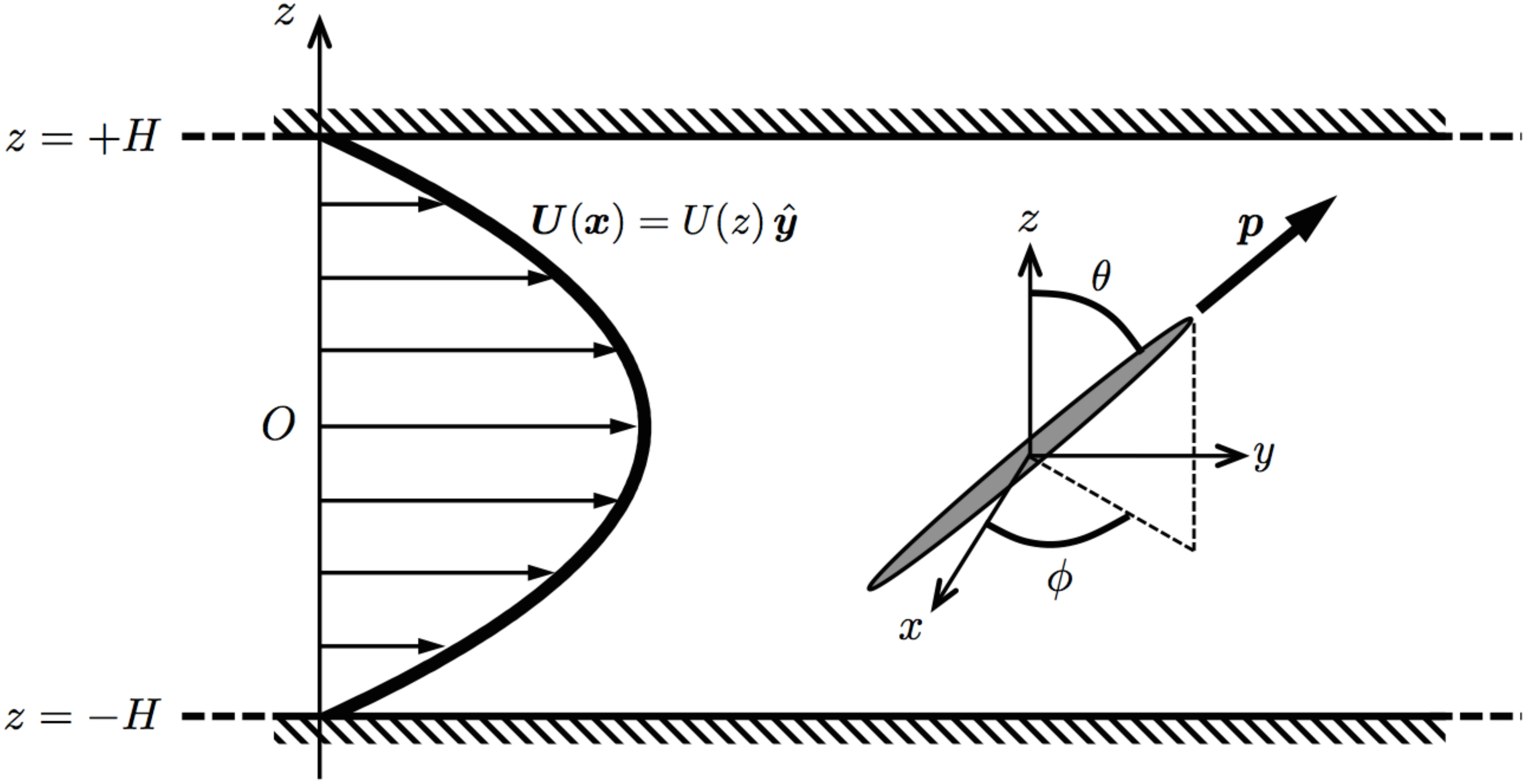}\vspace{0.2cm}
\caption{Problem definition: a dilute suspension of slender active particles with positions $\v{x}=(x,y,z)$ and orientations $\v{p}=(\sin\theta\cos\phi,\sin\theta\sin\phi,\cos\theta)$ is confined between two parallel flat plates ($z=\pm H$) and subject to an imposed pressure-driven parabolic flow.}\label{fig:definition}
\end{figure}

We analyze the dynamics in a dilute suspension of self-propelled slender particles confined between two parallel flat plates and placed in an externally imposed pressure-driven flow as illustrated in figure~\ref{fig:definition}. The channel half-width is denoted by $H$, and is assumed to be much greater than the characteristic length $L$ of the particles ($H/L\gg 1$), so that the finite size of the particles can be neglected. The external flow follows the parabolic Poiseuille profile 
\begin{equation}
\v{U}(\v{x})=U(z)\,\hat{\v{y}}=U_{m}\left[1-(z/H)^{2}\right]\hat{\v{y}},
\end{equation}
with maximum velocity $U_{m}$ at the centerline ($z=0$). The shear rate varies linearly with position $z$ across the channel:
\begin{equation}
S(z)=\frac{\mathrm{d} U}{\mathrm{d} z}=-\dot{\gamma}_{w}\frac{z}{H},
\end{equation}
where $\dot{\gamma}_{w}=2U_{m}/H$ is the maximum absolute shear rate attained at the walls ($z=\pm H$).

Following previous models for active suspensions \citep{SS2008a,SS2008b}, the configuration of the active particles  is captured by the probability distribution function $\Psi(\v{x},\v{p},t)$ of finding a particle at position $\v{x}=(x,y,z)$ with orientation $\v{p}=(\sin\theta\cos\phi,$ $\sin\theta\sin\phi,\cos\theta)$ at time $t$, where $\v{p}$ also defines the direction of swimming. Conservation of particles is expressed by the Smoluchowski equation \citep{Doi86}
\begin{equation}
\frac{\partial\Psi}{\partial t} +\bnabla_{x}\bcdot (\dot{\boldsymbol{x}}\,\Psi)+\bnabla_{p}\bcdot(\dot{\boldsymbol{p}}\,\Psi) =0, \label{eq:fp}
\end{equation}
where the translational flux velocity $\dot{\v{x}}$ captures self-propulsion with constant velocity $V_{s}$ in the direction of $\v{p}$, advection by the imposed flow, and center-of-mass diffusion with isotropic and constant diffusivity $d_{t}$:
\begin{equation}
\dot{\boldsymbol{x}}=V_{s}\,\boldsymbol{p}+\boldsymbol{U}(z)-d_{t} \bnabla_{{x}}\ln \Psi. \label{eq:xdot}
\end{equation}
Particle rotations are captured by the angular flux velocity $\dot{\v{p}}$, which includes contributions from the imposed flow via Jeffery's equation \citep{Jeffery22,Bretherton62}, and from rotational diffusion with diffusivity $d_{r}$: 
\begin{equation}
\dot{\boldsymbol{p}}=S(z)(\hat{\v{z}}\bcdot\v{p})(\t{I}-\v{pp})\bcdot\hat{\v{y}}-d_{r}\bnabla_{{p}}\ln \Psi. \label{eq:pdot}
\end{equation}
We have assumed that the particles have a high aspect ratio, a good approximation for common motile bacteria as well as many self-propelled catalytic micro-rods. Particle-particle hydrodynamic interactions have also been neglected based on the assumption of infinite dilution; such interactions could otherwise be included via an additional disturbance velocity in the expressions for $\dot{\v{x}}$ and $\dot{\v{p}}$ \citep{SS2008b}. As a result, we expect the distribution of particles to be uniform along the $x$ and $y$ directions, and at steady state the Smoluchowski equation (\ref{eq:fp}) for $\Psi(\v{x},\v{p},t)=\Psi(z,\v{p})$ then simplifies to
\begin{equation}
V_{s}\cos\theta\frac{\partial \Psi}{\partial z}-d_{t}\frac{\partial^{2}\Psi}{\partial z^{2}}+S(z)\bnabla_{p}\bcdot\left[\cos\theta (\t{I}-\v{pp})\bcdot\hat{\v{y}}\,\Psi\right]=d_{r}\nabla_{p}^{2}\Psi. \label{eq:goveq0}
\end{equation}
This equation simply expresses the balance of self-propulsion, translational diffusion, particle alignment by the imposed flow, and rotational diffusion. 

\textcolor{black}{In this work, we treat the translational and rotational diffusivities  $d_{t}$ and $d_{r}$ as independent constants, which could result from either Brownian motion or various athermal sources of noise \citep{Drescher11,Garcia11}. The athermal contribution to diffusion may arise due to tumbling or other fluctuations in the swimming actuation of motile micro-organisms, or from fluctuations in the chemical actuation mechanism of catalytic particles. In many active suspensions, such athermal fluctuations are in fact the dominant source of diffusion.}
\subsection{Boundary conditions \label{subsec:BC}}


In the continuum limit, the impenetrability of the channel walls is captured by prescribing that the normal component of the translational flux be zero at both walls: 
\begin{equation}
\hat{\v{z}}\bcdot\dot{\v{x}}=0 \quad \mathrm{at} \quad z=\pm H. \label{eq:BC00}
\end{equation} 
Inserting equation (\ref{eq:xdot}) for the translational flux, this leads to a Robin boundary condition for the probability distribution function:
\begin{equation}
d_{t}\frac{\partial \Psi}{\partial z}=V_{s}\cos\theta\, \Psi  \quad \mathrm{at} \quad z=\pm H, \label{eq:BC0}
\end{equation}
expressing the balance of translational diffusion and self-propulsion in the wall-normal direction. Equation (\ref{eq:BC0}) implies that particles pointing towards a wall ($\cos\theta>0$ for the top wall at $z=+H$) incur a positive wall-normal gradient ($\partial \Psi/\partial z>0$), whereas particles pointing away from the wall ($\cos\theta<0$) incur a negative gradient. This suggests that sorting of orientations should occur and lead to a net polarization towards the walls, accompanied by near-wall accumulation. These effects will indeed be confirmed in \S \ref{sec:noflow}. \textcolor{black}{It is important to note that the boundary condition (\ref{eq:BC0}) requires that the wall-normal swimming flux be balanced by a diffusive flux. In the complete absence of translational diffusion ($d_{t} = 0$), the swimming flux can no longer be balanced at the wall: this singular limit, which is ill-posed in our mean-field theory, will not be addressed here. Note also that the balance of the wall-normal fluxes hints at a length scale of $\ell_{a}=d_{t}/V_{s}$ for wall accumulation, as we demonstrate more quantitatively below.  }

\textcolor{black}{Other types of boundary conditions have been considered in previous works. In particular, several studies have implemented the condition
\begin{equation}
\int_{\Omega}\hat{\v{z}}\bcdot\dot{\v{x}}\,\Psi\,\mathrm{d}\v{p}=0 \quad \mathrm{at} \quad z=\pm H, \label{eq:BC000}
\end{equation} 
where $\Omega$ denotes the unit sphere of orientation. Equation (\ref{eq:BC000}) captures the zeroth orientational moment of  (\ref{eq:BC0}) and is easily implemented numerically using a reflection condition on the distribution function. 
It was first used by \cite{Bearon11} in a two-dimensional model of suspensions of gyrotactic swimmers constrained to a planar domain. \cite{Ezhilan12} also imposed equation (\ref{eq:BC000}) in the case of a chemotactic active suspension confined to a thin liquid film, where the primary mechanism for accumulation was chemotaxis as opposed to kinematics. In the absence of external fields, however, this boundary condition allows for a uniform isotropic solution throughout the channel and is therefore unable to capture near-wall accumulation or  upstream swimming when a flow is imposed (see Appendix~A for more details). \cite{Kasyap14a} also considered chemotactic active suspensions in thin films but used a position/orientation decoupling approximation for the probability distribution function $\Psi\left(\v{x},\v{p},t\right)$, allowing them to derive a boundary condition for the number density field expressing the balance of the chemotactic and diffusive fluxes at the boundaries. To our knowledge, the only previously reported use of the boundary condition (\ref{eq:BC0}) for a confined active suspension was in the work of \cite{Elgeti13}, whose analysis was restricted to equilibrium distributions in the absence of flow and in the limits of narrow channels or weak propulsion.}

Finally, it should be kept in mind that the simple boundary condition (\ref{eq:BC0}) neglects the finite size of the particles and is therefore inaccurate very close to the walls, where steric exclusion prohibits certain particle configurations and should lead to a depletion layer as observed in experiments \citep{Takagi14}. The implications of steric exclusion are discussed further in Appendix~B, where a more detailed boundary condition is derived and enforced on the hypersurface separating allowed from forbidden configurations \citep{Nitsche90,Schiek95,Krochak10}. As we show there, the effects of steric exclusion are weak in wide channels ($H/L\gg 1$) such as the ones considered in this work. 

\subsection{Dimensional analysis and scaling}

Dimensional analysis of the governing equations reveals three dimensionless groups:
\begin{equation}
Pe_{s}=\frac{V_{s}}{2d_{r}H}, \quad Pe_{f}=\frac{\dot{\gamma}_{w}}{d_{r}}, \quad \Lambda=\frac{d_{t}d_{r}}{V_{s}^{2}}.
\end{equation}
The first parameter $Pe_{s}$, or swimming P\'eclet number, can be interpreted as the ratio of the characteristic timescale for a particle to lose memory of its orientation due to rotational diffusion over the time it takes it to swim across the channel width. Equivalently, it is also the ratio of the persistence length of particle trajectories \textcolor{black}{($\ell_{p} = V_{s}/d_{r}$)} over the channel width \textcolor{black}{($2H$)}. The second parameter $Pe_{f}$, or flow P\'eclet number, compares the same diffusive timescale to the characteristic time for a particle to align under the imposed velocity gradient. The third parameter $\Lambda$ relates the translational and rotational diffusivities \textcolor{black}{to the swimming speed and is a fixed constant for a given particle type. It can be interpreted as an inverse measure of the strength of propulsion of a swimmer with respect to fluctuations, and the limits of $\Lambda \to 0$ and  $\Lambda \to \infty$ describe the strong and  weak propulsion cases, respectively. When $\Lambda$ is held constant, $Pe_{s}$ also reduces to an inverse measure of confinement, with $Pe_{s} \to 0$ and $Pe_{s} \to \infty$ describing the limits of weak and strong confinement, respectively.} 

In the following, we nondimensionalize the governing equations using the characteristic time, length and velocity scales
\begin{equation}
t_{c}=d_{r}^{-1}, \quad \ell_{c}=H, \quad v_{c}=H d_{r},
\end{equation}
and also normalize the distribution function $\Psi$ by the mean number density $n$ defined as
\begin{equation}
n=\frac{1}{2H}\int_{-H}^{H}\int_{\Omega}\Psi(z,\v{p})\,\mathrm{d}\v{p}\,\mathrm{d}z. \label{eq:psinorm}
\end{equation}
After nondimensionalization, the conservation equation (\ref{eq:goveq0}) becomes
\begin{equation}
Pe_{s}\cos\theta \frac{\partial \Psi}{\partial z}- \textcolor{black}{2\Lambda}Pe_{s}^{2}\frac{\partial^{2}\Psi}{\partial z^{2}}+\frac{Pe_{f}}{2}\,S(z)\bnabla_{p}\bcdot\left[\cos\theta (\t{I}-\v{pp})\bcdot\hat{\v{y}}\,\Psi\right]=\frac{1}{2}\nabla_{p}^{2}\Psi, \label{eq:goveq1}
\end{equation}
where the dimensionless shear rate profile is simply $S(z)=-z$. The boundary condition (\ref{eq:BC0}) also becomes
\begin{equation}
\frac{\partial \Psi}{\partial z}=
\textcolor{black}{\frac{1}{2\Lambda Pe_{s}}}\cos\theta\,\Psi \quad \mathrm{at} \quad z=\pm 1. \label{eq:BC1}
\end{equation}
\textcolor{black}{Note that the choice of $H$ for the characteristic length scale is convenient as it sets the positions of the boundaries to $z=\pm1$ in the dimensionless system. However, we will see below that alternate length scales are more judiciously chosen in certain limits due to the presence of boundary layers.}

\subsection{\textcolor{black}{Orientational m}oment equations\label{sec:moments}}

Equation (\ref{eq:goveq1}), together with boundary condition (\ref{eq:BC1}), cannot be solved analytically \textcolor{black}{in general.} While a numerical solution is possible as we show below, analytical progress can still be made in terms of orientational moments of the distribution function \citep{SS2013}. More precisely, we introduce the zeroth, first, and second moments of $\Psi(z,\v{p})$ as
\begin{align}
c(z)=\langle 1 \rangle, \quad \v{m}(z)=\langle \v{p}\rangle, \quad \t{D}(z)=\langle \v{pp}-\t{I}/3\rangle, \label{eq:momentsdef}
\end{align}
where the brackets $\langle\cdot\rangle$ denote the orientational average
\begin{equation}
\langle h(\v{p})\rangle=\int_{\Omega}h(\v{p})\Psi(z,\v{p})\,\mathrm{d}\v{p}.
\end{equation}
The zeroth moment $c(z)$ corresponds to the local concentration of particles. The next two moments are directly related to the polarization vector $\v{P}(z)$ and to the nematic order parameter tensor $\t{Q}(z)$ commonly used in the description of liquid-crystalline systems \citep{Marchetti13} as
\begin{equation}
\v{m}(z)=c(z)\v{P}(z), \quad \t{D}(z)=c(z)\t{Q}(z).
\end{equation}
Knowledge of these as well as higher moments also allows one to recover the full distribution function as
\begin{equation}
\Psi(z,\v{p})=\frac{1}{4\upi}\,c(z)+\frac{3}{4\upi}\,\v{p}\bcdot\v{m}(z)+\frac{15}{8\upi}\,\v{pp}\boldsymbol{:}\t{D}(z)+..., \label{eq:moments}
\end{equation}
which can also be interpreted as a spectral expansion of $\Psi(z,\v{p})$ on the basis of spherical harmonics. Near isotropy this expansion converges rapidly, which justifies truncation after a few terms. If only the first three terms corresponding to $c$, $\v{m}$ and $\t{D}$ are retained, a closed system of equations can be derived for these variables by taking moments of the conservation equation (\ref{eq:goveq1}) \citep{Baskaran09,SS2013}.

In the problem of interest to us here, symmetries dictate that the only non-zero components of $\v{m}$ and $\t{D}$ are $m_{z}$ and $D_{zz}=-2D_{xx}=-2D_{yy}$ in the absence of flow. When a flow is applied in the $y$ direction, $m_{y}$ and $D_{yz}=D_{zy}$ are also expected to become non-zero, and $D_{yy}$ need no longer be equal to $D_{xx}$. The governing equations for these variables can be obtained as\vspace{-0.1cm}
\begin{align}
Pe_{s} \frac{\mathrm{d} m_{z}}{\mathrm{d}z}-\textcolor{black}{2\Lambda}Pe_{s}^{2}\frac{\mathrm{d}^{2}c}{\mathrm{d}z^{2}}&=0, \label{eq:ceq} \\
Pe_{s}\frac{\mathrm{d}D_{zz}}{\mathrm{d}z}-\textcolor{black}{2\Lambda}Pe_{s}^{2}\frac{\mathrm{d}^{2}m_{z}}{\mathrm{d}z^{2}}+\textcolor{black}{\left(\frac{1}{6\Lambda}+1\right)}m_{z}&=-\frac{1}{10}Pe_{f}S(z)m_{y},  \\
Pe_{s}\frac{\mathrm{d}D_{yz}}{\mathrm{d}z}-\textcolor{black}{2\Lambda}Pe_{s}^{2}\frac{\mathrm{d}^{2}m_{y}}{\mathrm{d}z^{2}}+m_{y}&=\frac{2}{5}Pe_{f}S(z)m_{z}, \\
\frac{4}{15}Pe_{s}\frac{\mathrm{d}m_{z}}{\mathrm{d}z}-\textcolor{black}{2\Lambda}Pe_{s}^{2}\frac{\mathrm{d}^{2}D_{zz}}{\mathrm{d}z^{2}}+3D_{zz}&=\frac{4}{7}Pe_{f}S(z)D_{yz},\label{eq:Dzzeq} \\
-\frac{2}{15}Pe_{s}\frac{\mathrm{d}m_{z}}{\mathrm{d}z}-\textcolor{black}{2\Lambda}Pe_{s}^{2}\frac{\mathrm{d}^{2}D_{yy}}{\mathrm{d}z^{2}}+3D_{yy}&=-\frac{3}{7}Pe_{f}S(z)D_{yz},\label{eq:Dyyeq} \\
\frac{1}{5}Pe_{s}\frac{\mathrm{d}m_{y}}{\mathrm{d}z}-\textcolor{black}{2\Lambda}Pe_{s}^{2}\frac{\mathrm{d}^{2}D_{yz}}{\mathrm{d}z^{2}}+3D_{y\textcolor{black}{z}}&=Pe_{f}S(z)\left(\frac{1}{10}c+\frac{5}{14}D_{zz}-\frac{2}{7}D_{yy}\right). \label{eq:Dyzeq}\vspace{-0.1cm}
\end{align}
No equation is needed for $D_{xx}$, which can simply be deduced from $D_{yy}$ and $D_{zz}$ using the tracelessness of $\t{D}$.
In each of these equations, the first term on the left-hand side arises due to self-propulsion, the second term captures translational diffusion, and the third term rotational diffusion. Terms on the right-hand side arise from the externally applied pressure-driven flow and vanish in the absence of flow ($Pe_{f}=0$). 

Boundary conditions for these variables are also readily obtained by taking moments of equation (\ref{eq:BC1}), yielding\vspace{-0.1cm}
\begin{align}
&\frac{\mathrm{d}c}{\mathrm{d}z}=\textcolor{black}{\frac{1}{2\Lambda Pe_{s}}}m_{z}, \label{eq:cBC}\\
&\frac{\mathrm{d}m_{z}}{\mathrm{d}z}=\textcolor{black}{\frac{1}{2\Lambda Pe_{s}}}\left(D_{zz}+\frac{1}{3}c\right), \quad \frac{\mathrm{d}m_{y}}{\mathrm{d}z}=\textcolor{black}{\frac{1}{2\Lambda Pe_{s}}}D_{yz}, \\
&\frac{\mathrm{d}D_{zz}}{\mathrm{d}z}=\textcolor{black}{\frac{2}{15\Lambda Pe_{s}}}m_{z}, \quad \frac{\mathrm{d}D_{yy}}{\mathrm{d}z}=-\textcolor{black}{\frac{1}{15\Lambda Pe_{s}}}m_{z}, \quad \frac{\mathrm{d}D_{yz}}{\mathrm{d}z}=\textcolor{black}{\frac{1}{10\Lambda Pe_{s}}}m_{y}, \label{eq:DBC}\vspace{-0.1cm}
\end{align}
all to be enforced at $z=\pm 1$. For symmetry reasons, we  expect $c$, $m_{y}$, $D_{yy}$, $D_{zz}$ to be even functions of $z$, whereas $m_{z}$ and $D_{yz}$ are expected to be odd functions. \textcolor{black}{While we consider rotational diffusion as the only orientation decorrelation mechanism in this work, all the derivations shown here can be easily modified to account for run-and-tumble dynamics instead by modifying numerical prefactors in the third terms on the left-hand sides of equations~(\ref{eq:Dzzeq})--(\ref{eq:Dyzeq}).}

Integrating equation (\ref{eq:ceq}) and making use of the boundary condition (\ref{eq:cBC}) easily shows that (\ref{eq:ceq}) can be replaced by\vspace{-0.1cm}
\begin{equation}
m_{z}-\textcolor{black}{2\Lambda Pe_{s}}\frac{\mathrm{d}c}{\mathrm{d}z}=0 \label{eq:cmzrel}\vspace{-0.1cm}
\end{equation}
at every point in the channel, underlining the direct relation between transverse polarization and concentration gradients. We also note that the normalization condition (\ref{eq:psinorm}) on the distribution function translates into an integral condition on the concentration field expressing conservation of the total particle number:\vspace{-0.1cm}
\begin{equation}
\int_{-1}^{1}c(z)\,\mathrm{d}z=2. \label{eq:cnorm}\vspace{-0.1cm}
\end{equation}
As we discuss next, solution of the system (\ref{eq:ceq})--(\ref{eq:Dyzeq}) subject to the boundary conditions (\ref{eq:cBC})--(\ref{eq:DBC}) and to the integral constraint (\ref{eq:cnorm}) is possible under certain assumptions, and provides results that are in excellent quantitative agreement with the full numerical solution of the Smoluchowski equation (\ref{eq:goveq1}) over a wide range of values of the P\'eclet numbers. 

\section{Equilibrium distributions in the absence of flow\label{sec:noflow}}

We first analyze the case of no external flow ($Pe_{f}=0$), where we expect the boundary condition (\ref{eq:BC1}) to lead to near-wall accumulation and polarization as a result of self-propulsion. In this case, the full governing equation (\ref{eq:goveq1}) simplifies to 
\begin{equation}
Pe_{s}\left(\cos\theta\frac{\partial \Psi}{\partial z}-\textcolor{black}{2\Lambda}Pe_s\frac{\partial^{2}\Psi}{\partial z^{2}}\right)=\frac{1}{2}\nabla_{p}^{2}\Psi, \label{eq:goveq2}
\end{equation}
subject to condition (\ref{eq:BC1}) at the walls. 
\textcolor{black}{We note some interesting mathematical properties of these equations.  First, taking the cross-sectional average of equation (\ref{eq:goveq2}) yields
\begin{equation}
\nabla_{p}^{2}\left(\int_{-1}^{1} \Psi \,\mathrm{d}z\right) = 0,
\end{equation}
which implies that the gap-averaged orientation distribution is isotropic in the absence of flow. Using the conservation constraint (\ref{eq:cnorm}), we obtain
\begin{equation}
\int_{-1}^{1} \Psi \,\mathrm{d}z= \frac{1}{2\upi},
\end{equation}
which also implies that the first and higher-order moments all average to zero across the channel width when there is no flow.}

It is also easily seen that the uniform and isotropic distribution $\Psi=1/4\upi$ is an exact solution of equation (\ref{eq:goveq2}) \textcolor{black}{for all parameter values}, though it violates the boundary condition (\ref{eq:BC1}) when $\Lambda \neq \infty$. 
\textcolor{black}{Inspection of the equations shows that, in the limit of $\Lambda Pe_{s} = d_{t}/2V_{s}\rightarrow 0$, there is a loss of the higher derivative in both the governing equation and the boundary condition. This singular limit suggests the existence of an accumulation layer near the channel walls where the distribution departs from the uniform isotropic state. Inside this boundary layer, the effects of self-propulsion must be balanced by translational diffusion, notwithstanding the small value of $\Lambda Pe_{s}$.}  Rescaling the governing equation inside the boundary layer, however, does not lead to analytical simplifications \textcolor{black}{for finite $\Lambda$}, so we turn to the simplified moment equations for further characterization of particle distributions near the walls in \S \ref{sec:finite_L}, where a simple analytical solution is derived together with a scaling for the thickness of the accumulation layer. We then describe how the limits of strong and weak propulsion can be addressed using asymptotic expansions in \S \ref{sec:small_L} and \S\ref{sec:large_L}.

\subsection{Theory based on moment equations \label{sec:finite_L}}

In the absence of flow,  the moment equations derived in \S \ref{sec:moments} only involve $c$, $m_{z}$ and $D_{zz}$, and simplify to:
\begin{align}
m_{z}-\textcolor{black}{2\Lambda}Pe_{s}\frac{\mathrm{d}c}{\mathrm{d}z}=0,  \label{eq:ceq1}\\
Pe_{s}\frac{\mathrm{d}D_{zz}}{\mathrm{d}z}-\textcolor{black}{2\Lambda}Pe_{s}^{2}\frac{\mathrm{d}^{2}m_{z}}{\mathrm{d}z^{2}}+\textcolor{black}{\left(\frac{1}{6\Lambda}+1\right)}m_{z}=0, \label{eq:mzeq1} \\
\frac{4}{15}Pe_{s}\frac{\mathrm{d}m_{z}}{\mathrm{d}z}-\textcolor{black}{2\Lambda}Pe_{s}^{2}\frac{\mathrm{d}^{2}D_{zz}}{\mathrm{d}z^{2}}+3D_{zz}=0, \label{eq:Dzzeq1}
\end{align}
subject to the integral constraint (\ref{eq:cnorm}) and to the boundary conditions
\begin{equation}
\frac{\mathrm{d}m_{z}}{\mathrm{d}z}=\textcolor{black}{\frac{1}{2\Lambda Pe_{s}}}\left(D_{zz}+\frac{1}{3}c\right), \quad \frac{\mathrm{d}D_{zz}}{\mathrm{d}z}=\textcolor{black}{\frac{2}{15\Lambda Pe_{s}}}m_{z}\quad \mathrm{at}\,\,\, z=\pm1. \label{eq:BC2}
\end{equation}

Using this set of equations, we first proceed to derive a relation between the values of the concentration and wall-normal polarization at the boundaries. First, we integrate equation (\ref{eq:Dzzeq1}) across the channel width and use the second boundary condition in (\ref{eq:BC2}) to arrive at
\begin{equation}
\int_{-1}^{1}D_{zz}(z)\,\mathrm{d}z=0. \label{eq:Dzzint}
\end{equation}
Now, combining equations (\ref{eq:ceq1}) and (\ref{eq:mzeq1}), integrating from $z$ to $1$ and making use of the first boundary condition gives
\begin{equation}
D_{zz}-\textcolor{black}{2\Lambda Pe_{s}}\frac{\mathrm{d}m_{z}}{\mathrm{d}z}+\textcolor{black}{\frac{6\Lambda + 1}{3}}c=\textcolor{black}{2\Lambda}\,c(1).
\end{equation}
This relation can be integrated once more across the channel width. Using condition (\ref{eq:Dzzint}) together with the parity properties of $c$ and $m_{z}$, this simplifies to
\begin{equation}
c(\pm 1)=\textcolor{black}{\left(1+\frac{1}{6\Lambda}\right)}\mp Pe_{s}m_{z}(\pm1), \label{eq:cmz}
\end{equation}
providing a simple relation between concentration and polarization at the walls. Inserting this relation into the first condition in (\ref{eq:BC2}) yields a new set of boundary conditions that does not involve the concentration:
\begin{equation}
\frac{\mathrm{d}m_{z}}{\mathrm{d}z}=\textcolor{black}{\frac{1}{2\Lambda Pe_{s}}}\left(D_{zz}+\textcolor{black}{\frac{6\Lambda + 1}{18\Lambda}}\right)\mp \textcolor{black}{\frac{1}{6\Lambda}}m_{z}, \quad \frac{\mathrm{d}D_{zz}}{\mathrm{d}z}=\textcolor{black}{\frac{2}{15\Lambda Pe_{s}}}m_{z}\quad \mathrm{at}\,\,\, z=\pm1.
\end{equation}
Equations (\ref{eq:mzeq1})--(\ref{eq:Dzzeq1}), together with these boundary conditions, form a coupled system of second-order linear ordinary differential equations for $m_{z}$ and $D_{zz}$ that can be solved analytically. Once these variables are known, the concentration profile is easily obtained from the polarization by integration of (\ref{eq:ceq1}) along with condition (\ref{eq:cmz}).

Solving these equations yields complicated expressions for $c$, $m_{z}$ and $D_{zz}$ that are omitted here for brevity. The profiles, which are illustrated in figure~\ref{fig:noflow_Pes} and will be discussed in more detail below, reveal one important finding: while a significant wall-normal polarization exists in the near-wall region, nematic alignment is relatively weak throughout the channel \textcolor{black}{for $\Lambda \gtrsim 0.1$}. This suggests seeking a yet simpler solution that neglects nematic order altogether. If the moment expansion (\ref{eq:moments}) is truncated after two terms, the equations for $c$ and $m_{z}$ simplify to
\begin{equation}
m_{z}-\textcolor{black}{2\Lambda Pe_{s}}\frac{\mathrm{d}c}{\mathrm{d}z}=0, \quad -\textcolor{black}{2\Lambda}{Pe}_{s}^{2}\frac{\mathrm{d}^{2}m_{z}}{\mathrm{d}z^{2}}+\textcolor{black}{\left(\frac{1}{6\Lambda}+1\right)}m_{z}=0,
\end{equation}
subject to the conditions
\begin{equation}
\frac{\mathrm{d}m_{z}}{\mathrm{d}z}=\frac{c}{\textcolor{black}{6\Lambda} Pe_{s}}\,\,\,\,\, \mathrm{at}\,\,\,\, z=\pm 1, \quad \mathrm{and} \quad \int_{-1}^{1}c(z)\,\mathrm{d}z=2.
\end{equation}
Solving these equations is straightforward and  provides elegant expressions for the concentration and polarization profiles:
\begin{align}
c(z)&=\frac{\textcolor{black}{B}\left[\textcolor{black}{6\Lambda}\cosh \textcolor{black}{B}+\cosh \textcolor{black}{Bz}\right]}{\textcolor{black}{6\Lambda B}\cosh \textcolor{black}{B} + \sinh \textcolor{black}{B}}, \label{eq:ceq2M}\\
m_{z}(z)&=\frac{\textcolor{black}{6\Lambda Pe_{s} B^{2}}\sinh \textcolor{black}{Bz}}{3\left(\textcolor{black}{6\Lambda B}\cosh \textcolor{black}{B} + \sinh \textcolor{black}{B}\right)},
\end{align}
where 
\begin{align}
\textcolor{black}{B^{-1} = \Lambda Pe_{s}\sqrt{\frac{12}{1+6\Lambda}}} 
\end{align}
\textcolor{black}{defines the dimensionless decay length of the excess concentration at the walls. In dimensional terms,  this decay length is given by $B^{-1}H = \ell_{a}\sqrt{3/(1+6\Lambda)}$ where $\ell_{a}=d_{t}/V_{s}$. In the limit of strong propulsion ($\Lambda \ll 1$), it simplifies to $\sqrt{3}\,\ell_{a}$. In the limit of weak propulsion ($\Lambda \gg 1$), it becomes $\ell_{d}/\sqrt{2}$ where $\ell_{d} = \sqrt{d_{t}/d_{r}}$ is a purely diffusive length scale. For Brownian particles, $\ell_{d}$ is typically of the order of the particle size $L$, though this may not be the case for active particles subject to athermal sources of noise.  
Next, we focus more precisely on  these two limits by rescaling the governing equations with the appropriate scales identified here.} 

\subsection{\textcolor{black}{Strong propulsion limit: $\Lambda \rightarrow 0$}\label{sec:small_L}}

\textcolor{black}{In the limit of small $\Lambda$, the above discussion suggests rescaling the Smoluchowski equation using the accumulation length scale $\ell_{a}$, yielding
\begin{equation}
\cos\theta\frac{\partial \Psi}{\partial z}-\frac{\partial^{2}\Psi}{\partial z^{2}}= \Lambda\nabla_{p}^{2}\Psi, \label{eq:goveq2b}
\end{equation}
subject to the boundary condition
\begin{equation}
\frac{\partial \Psi}{\partial z}=\cos\theta\,\Psi \quad \mathrm{at} \quad z=\pm \textcolor{black}{H^{*}}. \label{eq:BC1b}
\end{equation}
Here, $H^{*} = (2\Lambda Pe_{s})^{-1}$ is the channel half-height rescaled by the accumulation length scale $\ell_{a}$. \textcolor{black}{The gap-averaged isotropy constraint is now expressed as}\vspace{-0.1cm}
\begin{equation}
\int_{-H^{*}}^{H^{*}}\Psi \mathrm{d}z = \frac{H^{*}}{2\upi}. \label{eq:gavIso}\vspace{-0.1cm}
\end{equation}
The leading-order solution corresponding to $\Lambda =0$, which was previously obtained by \cite{Elgeti13}, is written
\begin{equation}
\Psi^{(0)}(z,\theta) = \frac{H^{*}\cos\theta}{4\upi \sinh \left(H^{*}\cos\theta\right)}\exp\left(z\cos\theta\right),
\end{equation}
and it is easily seen that it satisfies zero wall-normal flux pointwise throughout the channel. In particular, it shows that wall accumulation is possible even in the absence of rotational diffusion and is simply a result of a coupling between self-propulsion, translational diffusion and confinement. This solution can then be corrected to order $O(\Lambda)$ by solving the first-order inhomogeneous equation 
\begin{equation}
\cos\theta\,\Psi^{(1)}(z,\theta) - \frac{\partial \Psi^{(1)}}{\partial z} = \nabla_{p}^{2}\int_{-H^{*}}^{z}\Psi^{(0)}(z,\theta)\,\mathrm{d}z.
\end{equation}
subject to boundary condition (\ref{eq:BC1b}). An exact analytical solution to this equation can again be obtained but is cumbersome and omitted  here for brevity. }

\subsection{\textcolor{black}{Weak propulsion limit: $\Lambda \rightarrow \infty$}\label{sec:large_L}}

\textcolor{black}{In the limit of large $\Lambda$, the Smoluchowski equation is rescaled using the diffusive length scale $\ell_{d}$ as
\begin{equation}
\frac{1}{\sqrt{\Lambda}}\cos\theta\frac{\partial \Psi}{\partial z}-\frac{\partial^{2}\Psi}{\partial z^{2}}= \nabla_{p}^{2}\Psi, \label{eq:goveq2b}
\end{equation}
subject to
\begin{equation}
\frac{\partial \Psi}{\partial z}=\frac{1}{\sqrt{\Lambda}}\cos\theta\,\Psi \quad \mathrm{at} \quad z=\pm \textcolor{black}{H^{\dagger}}. \label{eq:BC1b}
\end{equation}
\noindent where $H^{\dagger} = (2\sqrt{\Lambda} Pe_{s})^{-1}$. The leading-order solution in the limit of $\Lambda \to \infty$ is uniform and isotropic and corresponds to the case of a passive particle: $\Psi^{(0)}(z,\theta)=1/4\upi$. It can be corrected asymptotically using a regular perturbation expansion in powers of $\Lambda^{-1/2}$: 
\begin{equation}
\Psi\left(z,\theta\right) = \Psi^{(0)}\left(z,\theta\right) + \Lambda^{-1/2}\,\Psi^{(1)}\left(z,\theta\right) + \Lambda^{-1}\,\Psi^{(2)}\left(z,\theta\right) + ... \label{eq:weakexp}
\end{equation}
Recursively solving for higher-order terms yields
\begin{align}
\Psi^{(1)}\left(z,\theta\right) &= \frac{3}{4\upi\sqrt{2}\cosh\left(\sqrt{2}H^{\dagger}\right)}\sinh(\sqrt{2}z) \cos \theta, \\
\Psi^{(2)}\left(z,\theta\right) &= \left[-\frac{1}{15}\frac{\cosh \sqrt{2}z}{\cosh (\sqrt{2}H^{\dagger})} + \frac{\tanh(\sqrt{2}H^{\dagger})}{5\sqrt{3}}\frac{\cosh(\sqrt{6}z)}{\sinh(\sqrt{6}H^{\dagger})}\right]\left(\cos^{2}\theta - \frac{1}{3}\right),
\end{align}
which both satisfy the appropriate boundary conditions. Quite remarkably, it can be seen that successive terms in the expansion (\ref{eq:weakexp}) correspond to successive orientational moments of the distribution function in equation (\ref{eq:moments}), with $\Psi^{(1)}$ and $\Psi^{(2)}$ describing the polarization and nematic order, respectively. }

\subsection{Numerical results and discussion\label{sec:numres}}

\begin{figure}
\centering\vspace{0.2cm}
\includegraphics{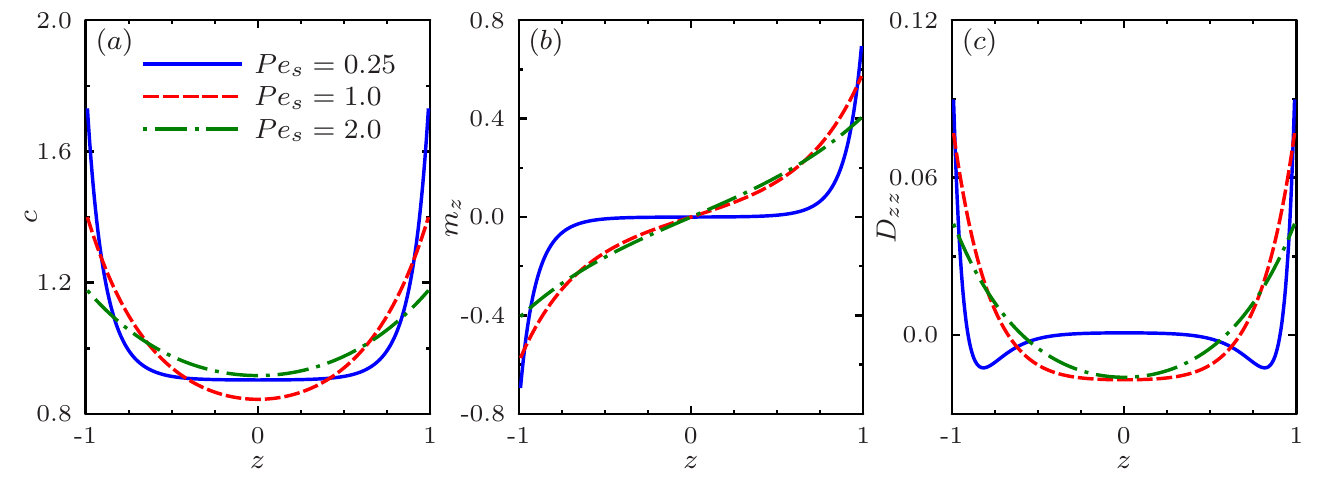}
\caption{(Color online) Equilibrium distributions in the absence of flow and for various swimming P\'eclet numbers $Pe_{s}$ (with $\Lambda = 1/6$), obtained by numerical solution of equation (\ref{eq:goveq1}) using finite volumes: (\textit{a}) concentration $c$, (\textit{b}) wall-normal polarization $m_{z}$, and (\textit{c}) wall-normal nematic order parameter $D_{zz}$. }\label{fig:noflow_Pes}
\end{figure}

\begin{figure}
\centering\vspace{0.2cm}
\includegraphics{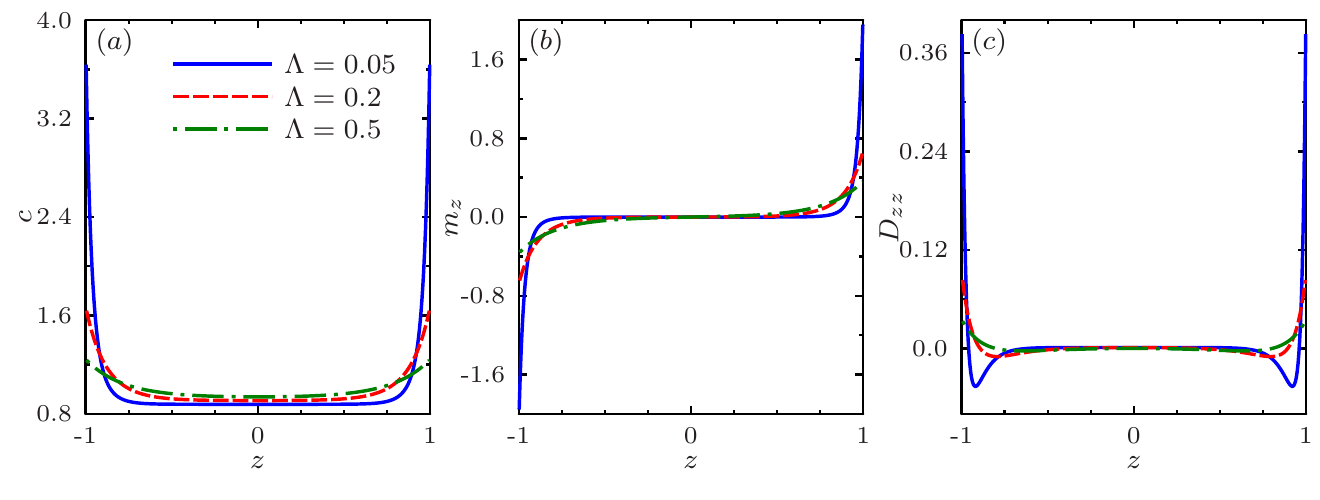}
\caption{(Color online) Equilibrium distributions in the absence of flow and for various values of $\Lambda$ (with $Pe_{s} = 0.25$), obtained by numerical solution of equation (\ref{eq:goveq1}) using finite volumes: (\textit{a}) concentration $c$, (\textit{b}) wall-normal polarization $m_{z}$, and (\textit{c}) wall-normal nematic order parameter $D_{zz}$. Solutions based on moment equations are nearly identical, as illustrated in figure~\ref{fig:noflow_error}.}\label{fig:noflow_Lambda}
\end{figure}

Figure~\ref{fig:noflow_Pes} shows the full numerical solution for the concentration $c$,  wall-normal polarization $m_z$ and nematic order parameter $D_{zz}$ obtained by finite-volume solution of the Smoluchowski equation (\ref{eq:goveq1}) as described in Appendix~B. \textcolor{black}{Here, we fix the value of $\Lambda$ and focus on the effect of $Pe_{s}$, which an inverse measure of confinement}. The concentration profiles shown in figure~\ref{fig:noflow_Pes}(\textit{a})  exhibit significant accumulation of particles near the boundaries, especially at low values of $Pe_{s}$. As anticipated, this accumulation is accompanied by polarization towards the boundaries as a direct consequence of the boundary condition (\ref{eq:cBC}), as well as by a weak nematic alignment. As $Pe_{s}$ increases, the spatial heterogeneity and anisotropy near the walls progressively extend through the entire channel as the two boundary layers thicken and eventually merge. Further increase in the swimming P\'eclet number leads to a flattening of the profiles, which is especially significant when $Pe_{s}>1$. This flattening is a direct consequence of the scaling of translational diffusion with $Pe_{s}^{2}$ in equation (\ref{eq:goveq1}), causing it to overwhelm self-propulsion which scales with $Pe_{s}$. \textcolor{black}{The influence of $\Lambda$ is illustrated in figure~\ref{fig:noflow_Lambda}, where it is seen to be similar to that of $Pe_{s}$: increasing $\Lambda$ leads to a thickening of the boundary layers and flattening of the concentration profiles, again due to the scaling of translational diffusion with $\Lambda$ in equation (\ref{eq:goveq1}).}

\begin{figure}
\centering\vspace{0.2cm}
\includegraphics{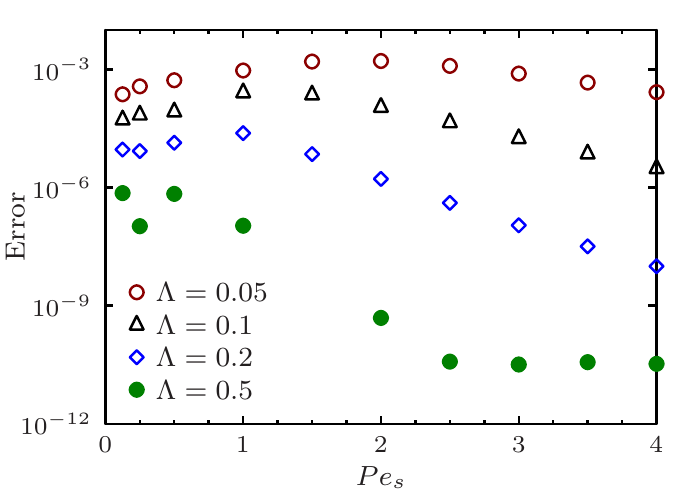}
\caption{(Color online) The relative rms error for the concentration between the finite-volume  solution and the two-moment analytical solution (\ref{eq:ceq2M}) for different values of $\Lambda$. Solutions based on moment equations are nearly identical to the finite-volume solution for sufficiently large values of $\Lambda$.}\label{fig:noflow_error}
\end{figure}

The finite-volume numerical solution of the full conservation equation (\ref{eq:goveq1}) is in excellent quantitative agreement with the two- and three-moment approximations derived previously, which are not shown in figure~\ref{fig:noflow_Pes} as they are nearly indistinguishable over the entire channel width \textcolor{black}{as long as $\Lambda \gtrsim 0.1$.} The rms error between the two-moment solution of equation (\ref{eq:ceq1}) and the finite-volume solution is indeed plotted in figure~\ref{fig:noflow_error}, where it remains below $10^{-3}$ for all values of $Pe_{s}$ considered here \textcolor{black}{when $\Lambda \gtrsim 0.1$}. This finding may seem quite surprising considering the strong approximation made when truncating expansion (\ref{eq:moments}) after only two terms, and strongly validates the use of approximate moment equations such as (\ref{eq:ceq})--(\ref{eq:Dyzeq}) when modeling active suspensions, at least in the absence of flow. \textcolor{black}{For very small values of $\Lambda$, however, nematic alignment at the walls becomes significant as seen in figure~\ref{fig:noflow_Lambda}(\textit{c}),  so that the nematic tensor can no longer be neglected and the two-moment solution loses its accuracy; in this case, the alternate expressions derived in the small $\Lambda$ limit in \S\ref{sec:small_L} can be used instead.} 

\begin{figure}
\centering\vspace{0.2cm}
\includegraphics{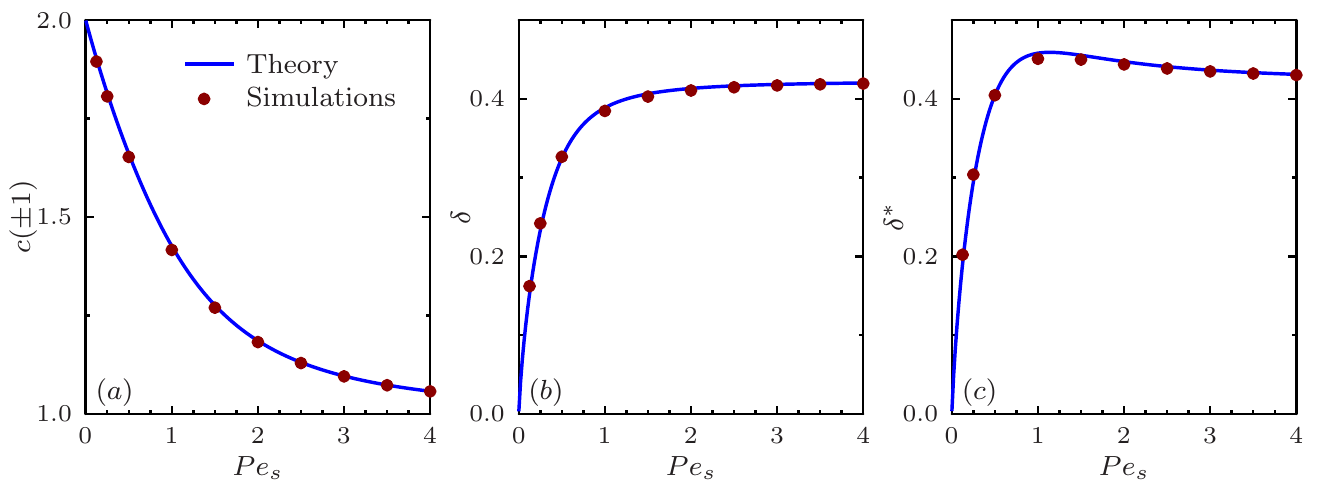}
\caption{(Color online) Wall accumulation in the absence of flow as a function of $Pe_{s}$ (at $\Lambda = 1/6$): (\textit{a}) concentration $c(\pm1)$ at the walls; (\textit{b}) boundary layer thickness $\delta$, defined as the distance from the wall where $c(1-\delta)=1$; (\textit{c}) fraction  $\delta^*$ of particles inside the boundary layer, defined as the integral of $c(z)$ over the boundary layer thickness. The solid line shows the theoretical prediction based on the two-moment solution (\ref{eq:ceq2M}), and symbols show full numerical results using finite volumes.}\label{fig:BLthickness}
\end{figure}

The influence of $Pe_{s}$ on wall accumulation is analyzed more quantitatively in figure~\ref{fig:BLthickness}, showing the values of the wall concentration $c(\pm1)$, the boundary layer thickness $\delta$ defined as the distance from the wall where $c(1-\delta)=1$, and the fraction $\delta^*$ of particles inside the boundary layer defined as 
\begin{equation}
\delta^{*}=\int_{1-\delta}^{1}c(z)\,\mathrm{d}z.
\end{equation}
Analytical expressions for these quantities can be derived from the two-moment solution (\ref{eq:ceq2M}). In particular, the boundary layer thickness is obtained as
\textcolor{black}{
\begin{equation}
\delta(Pe_{s}) = 1 - \frac{1}{B}\log\left\{\frac{\sinh B}{B} \pm \left[\left(\frac{\sinh B}{B}\right)^2 -1\right]^{1/2}\right\},
\end{equation}}which has the two limits
\begin{equation}
\lim_{Pe_{s}\rightarrow 0}\delta(Pe_{s})=0 \quad \mathrm{and} \quad \lim_{Pe_{s}\rightarrow \infty} \delta(Pe_{s})=1-\frac{1}{\sqrt{3}}.
\end{equation}
Similarly, the fraction of particles inside the boundary layer is given by
\textcolor{black}{
\begin{equation}
\delta^{*}(Pe_{s})=1- \frac{6\Lambda B\left(1-\delta\right) \cosh B + \sinh\left[B\left(1-\delta\right)\right]}{6\Lambda B \cosh B + \sinh B},
\end{equation}}and has the same limits as $\delta(Pe_{s})$ when $Pe_{s}\rightarrow 0$ and $\infty$. 

As shown in figure~\ref{fig:BLthickness}(\textit{a}), the wall concentration reaches its maximum  in the limit of $Pe_{s}\rightarrow 0$, and steadily decreases towards $1$ as $Pe_{s}$ increases due to the smoothing effect of translational diffusion. This is accompanied by an increase in the boundary layer thickness $\delta$, which asymptotes at high values of $Pe_{s}$. The fraction $\delta^*$ of particles near the walls shows a similar trend, but interestingly also exhibits a weak maximum for $Pe_{s}\approx 1.135$ when wall accumulation due to self-propulsion and translational diffusion are of similar magnitudes; at this value of $Pe_{s}$, $\delta^{*}\approx 0.46$ corresponding to nearly half the particles being trapped near the walls. As previously observed in figure~\ref{fig:noflow_error}, excellent agreement is obtained between the two-moment approximation and the numerical solution of the full governing equations. 

\section{Equilibrium distributions and transport in flow\label{sec:flow}}

\subsection{Weak-flow limit: regular asymptotic expansion\label{sec:weakflow}}

We now proceed to analyze the effects of an external pressure-driven flow, first focusing on the case of a weak flow for which $Pe_{f}\ll1$. \textcolor{black}{Since the parameter $\Lambda$ is fixed for a given type of swimmers, we keep it constant in the rest of the paper and focus on the effects of $Pe_{s}$ and $Pe_{f}$.} The form of the governing equations suggests seeking an approximate solution as a regular expansion of the moments of the distribution function in powers of $Pe_{f}$. The leading-order $O(Pe_{f}^{0})$ solution corresponding to the absence of flow was previously calculated in \S \ref{sec:noflow}. It is henceforth denoted by $c^{(0)}$, $\v{m}^{(0)}$, $\t{D}^{(0)}$, and we recall that $m_{y}^{(0)}=D_{yz}^{(0)}=0$. Inspection of the moment equations (\ref{eq:ceq})--(\ref{eq:Dyzeq}) reveals that the interaction of  the applied shear profile $S(z)$ with this leading-order solution perturbs $m_{y}$ and $D_{yz}$ at order $O(Pe_{f})$. On the other hand, $c$, $m_{z}$, $D_{zz}$ and $D_{yy}$ are only perturbed by the flow at order $O(Pe_{f}^{2})$ due to its interaction with $m_{y}$ and $D_{yz}$. Based on these observations, we expand the solution as
\begin{align}
c(z)&=c^{(0)}(z)+Pe_{f}^{2}\,c^{(2)}(z)+O(Pe_{f}^{3}), \\
m_{z}(z)&=m_{z}^{(0)}(z)+Pe_{f}^{2}m_{z}^{(2)}(z)+O(Pe_{f}^{3}), \\
D_{zz}(z)&=D_{zz}^{(0)}(z)+Pe_{f}^{2}D_{zz}^{(2)}(z)+O(Pe_{f}^{3}), \\
D_{yy}(z)&=D_{yy}^{(0)}(z)+Pe_{f}^{2}D_{yy}^{(2)}(z)+O(Pe_{f}^{3}), \\
m_{y}(z)&=Pe_{f}m_{y}^{(1)}(z)+O(Pe_{f}^{3}), \\
D_{yz}(z)&=Pe_{f}D_{yz}^{(1)}(z)+O(Pe_{f}^{3}).
\end{align}
We focus here on determining the leading-order corrections to $m_{y}$ and $D_{yz}$, which capture streamwise polarization and nematic alignment with the applied shear, respectively. The $O(Pe_{f})$ moment equations are written
\begin{align}
Pe_{s}\frac{\mathrm{d}D_{yz}^{(1)}}{\mathrm{d}z}-\textcolor{black}{2\Lambda}{Pe}_{s}^{2}\frac{\mathrm{d}^{2}m_{y}^{(1)}}{\mathrm{d}z^{2}}+m_{y}^{(1)}&=\frac{2}{5}S(z)m_{z}^{(0)}, \label{eq:my1eq}\\
\frac{Pe_{s}}{5}\frac{\mathrm{d}m_{y}^{(1)}}{\mathrm{d}z}-\textcolor{black}{2\Lambda}Pe_{s}^{2}\frac{\mathrm{d}^{2}D_{yz}^{(1)}}{\mathrm{d}z^{2}}+3D_{yz}^{(1)}&=S(z)\left(\frac{1}{10}c^{(0)}+\frac{1}{2}D_{zz}^{(0)}\right), \label{eq:Dyz1eq}
\end{align}
subject to boundary conditions
\begin{equation}
\frac{\mathrm{d}m_{y}^{(1)}}{\mathrm{d}z}=\textcolor{black}{\frac{1}{2\Lambda Pe_{s}}}D_{yz}^{(1)}, \quad \frac{\mathrm{d}D_{yz}^{(1)}}{\mathrm{d}z}=\textcolor{black}{\frac{1}{10\Lambda Pe_{s}}}m_{y}^{(1)} \quad \mathrm{at}\,\,\, z=\pm1. \label{eq:BCmy1}
\end{equation}
Note that the forcing terms on the right-hand sides of equations (\ref{eq:my1eq})--(\ref{eq:Dyz1eq}) are known and capture the interaction of the local shear rate $S(z)$ with the equilibrium distributions in the absence of flow.

\begin{figure}
\centering\vspace{0.2cm}
\includegraphics{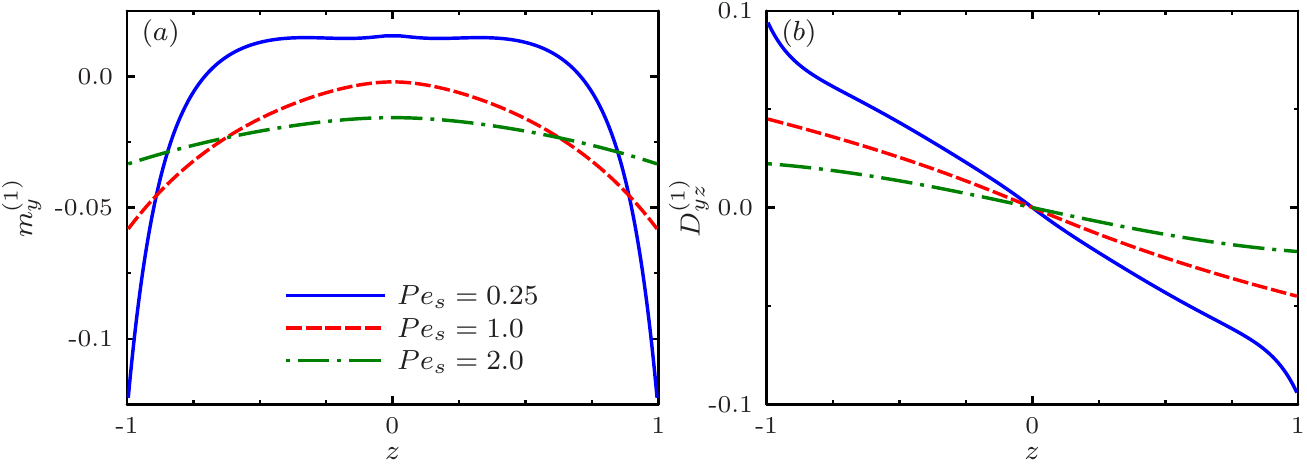}
\caption{(Color online) Effect of a weak applied flow: leading-order $O(Pe_{f})$ corrections of (\textit{a}) streamwise polarization $m_y$ and (\textit{b}) shear nematic alignment $D_{yz}$ for different values of the swimming P\'eclet number, obtained by numerical solution of equations (\ref{eq:my1eq})--(\ref{eq:BCmy1}).}\label{fig:weakflow}
\end{figure}

A numerical solution of equations (\ref{eq:my1eq})--(\ref{eq:BCmy1}) is plotted in figure~\ref{fig:weakflow} for different values of $Pe_{s}$. At low values of the swimming P\'eclet number, figure~\ref{fig:weakflow}(\textit{a}) shows an upstream polarization ($m_{y}<0$) near the boundaries, and a downstream polarization ($m_{y}>0$) near the center of the channel. The upstream polarization, which has previously been observed in both experiments and simulations and is at the origin of the well-known phenomenon of upstream swimming, is a simple and direct consequence of the shear rotation of the particles near the wall, which tend to point towards the walls in the absence of flow as explained in \S \ref{sec:noflow}. This interaction is encapsulated in the right-hand side in equation (\ref{eq:my1eq}). The downstream polarization near the centerline is a more subtle effect arising from self-propulsion through the first term on the left-hand side of (\ref{eq:my1eq}). As $Pe_{s}$ increases and the boundary layers thicken, upstream swimming becomes weaker near the boundaries due to the weaker wall-normal polarization there; however, $m_{y}$ is also observed to become negative across the entire channel due to the thickening of the polarized boundary layers into the bulk of the channel as previously shown in figure~\ref{fig:noflow_Pes}(\textit{b}).

The mean streamwise swimming velocity $\overline{V}_{y}$ of the active particles with respect to the imposed flow can be defined in terms of the polarization as
\begin{equation}
\overline{V}_{y}=\frac{1}{2}\int_{-1}^{1}Pe_{s}\,m_{y}(z)\,\mathrm{d}z=\frac{Pe_{s}Pe_{f}}{2}\int_{-1}^{1}m_{y}^{(1)}(z)\,\mathrm{d}z={Pe_{s}Pe_{f}}\,\overline{m}_{y}^{(1)}. \label{eq:upswim}
\end{equation}
An expression for $\overline{m}_{y}^{(1)}$ can be derived based on the moment equations. We first take the cross-sectional average of equation (\ref{eq:my1eq}) and use the first boundary condition to obtain
\begin{equation}
\overline{m}^{(1)}_{y}=-\textcolor{black}{\frac{1}{5}}\int_{-1}^{1}z\,m_{z}^{(0)}(z)\,\mathrm{d}z.
\end{equation}
Since $m_{z}^{(0)}$ is an odd function of $z$ with $m_{z}^{(0)}(z) \ge 0$ for $z \ge 0$, the integrand on the right-hand side is always positive across the channel, and therefore the mean upstream polarization is negative: $\overline{m}_{y}^{(1)} < 0$. This also implies that $\overline{V}_{y} < 0$, i.e., there is a net upstream flux of particles against the mean flow for all values of $\Lambda$ and $Pe_{s}$ in the weak flow limit. Using equation (\ref{eq:mzeq1}) for $m_{z}^{(0)}(z)$, we can rewrite the right-hand side as
\begin{equation}
\overline{m}_{y}^{(1)}=-\frac{1}{5\textcolor{black}{\left(\frac{1}{6\Lambda}+1\right)}}\left[\textcolor{black}{2\Lambda} Pe_{s}^{2}\int_{-1}^{1}z\,\frac{\mathrm{d}^{2}m_{z}^{(0)}}{\mathrm{d}z^{2}}\,\mathrm{d}z-Pe_{s}\int_{-1}^{1}z\,\frac{\mathrm{d}D_{zz}^{(0)}}{\mathrm{d}z}\,\mathrm{d}z\right].
\end{equation}
After integration by parts and application of the boundary condition on $m_{z}^{(0)}(z)$ together with equation (\ref{eq:Dzzint}), this simplifies to
\begin{equation}
\overline{m}_{y}^{(1)}=-\frac{\textcolor{black}{2}Pe_{s}}{15\textcolor{black}{\left(\frac{1}{6\Lambda}+1\right)}}\left[c^{(0)}(1)-\textcolor{black}{6\Lambda}Pe_{s}\,m_{z}^{(0)}(1)\right].
\end{equation}
Recalling that  $c^{(0)}(1)$ and $m_{z}^{(0)}(1)$ are related via equation (\ref{eq:cmz}), we obtain two expressions for the mean streamwise swimming velocity in terms of either the concentration or wall-normal polarization at the top wall in the absence of flow: 
\begin{equation}
\overline{V}_{y}=-\textcolor{black}{\frac{4\Lambda}{5}}Pe_{s}^{2}Pe_{f}\left[c^{(0)}(1)-1\right]=-\frac{2}{15}Pe_{s}^{2}Pe_{f}\left[1-\textcolor{black}{6\Lambda} Pe_{s}m_{z}^{(0)}(1)\right]. \label{eq:upvel}
\end{equation}
Since the concentration at the wall in the absence of flow always exceeds the mean when $Pe_{s}>0$,  equation (\ref{eq:upvel}) again confirms  that $\overline{V}_{y}<0$.
If we further make use of the simplified two-moment analytical solution (\ref{eq:ceq2M}) for the concentration profile, we arrive at a simple expression for the mean upstream velocity in terms of the swimming and flow P\'eclet numbers:
\begin{equation}
\overline{V}_{y}=-\textcolor{black}{\frac{4\Lambda}{5}Pe_{s}^{2}Pe_{f}\left[\frac{B \cosh B-\sinh B}{6\Lambda B\cosh B+\sinh B}\right]}.
\end{equation}
This simple analytical prediction for $\overline{V}_{y}$ will be tested against numerical simulations at arbitrary $Pe_{f}$ in \S \ref{sec:strongflow}, where it will be shown to provide an excellent estimate for the swimming flux up to $Pe_{f}\approx 2$. 

The effects of the external flow on nematic alignment are also illustrated in figure~\ref{fig:weakflow}(\textit{d}), where $D_{yz}$ is found to vary almost linearly across the channel width and has the same sign as the external shear rate profile $S(z)$. The right-hand side in equation (\ref{eq:Dyz1eq}) provides a simple explanation for these findings, where we see that shear nematic alignment results primarily from the interaction of the flow with the concentration profile and with the wall-normal nematic alignment. As $Pe_{s}$ increases, shear nematic alignment decreases due to the decrease in $c$ and $D_{zz}$ inside the boundary layers as seen in figures~\ref{fig:noflow_Pes}(\textit{a}) and (\textit{c}), and to self-propulsion through the first term on the left-hand side of equation (\ref{eq:Dyz1eq}). 

\subsection{Strong-flow limit: scaling analysis\label{sec:strongflow}}

As we shall see in \S \ref{sec:trapping} and figure~\ref{fig:arbitraryPec}, the regime of high flow P\'eclet number is also quite interesting as it can result in a depletion near the channel centerline surrounded by regions where particles become trapped. The thickness of this depletion region will be found to decrease with increasing flow strength, suggesting the presence of another boundary layer  near $z=0$ in the limit of $Pe_{f}\gg 1$. Insight into this regime can be gained by analyzing the behavior of the governing equation (\ref{eq:goveq1}) for $Pe_{f}\gg1$ and $Pe_{s}\ll 1$. If the swimming P\'eclet number is low, the wall boundary layers are very thin and have negligible impact on the dynamics in the bulk of the channel. Inspection of equation (\ref{eq:goveq1}) suggests that, in the outer region away from both the channel walls and the centerline, the dominant balance is between shear alignment and rotational diffusion:
\begin{equation}
\frac{Pe_{f}}{2}\,S(z)\bnabla_{p}\bcdot\left[\cos\theta (\t{I}-\v{pp})\bcdot\hat{\v{y}}\,\Psi\right]\approx \frac{1}{2}\nabla_{p}^{2}\Psi. 
\end{equation}
In this region, the concentration is expected to be nearly uniform, and the particle orientation distribution is primarily nematic as a result of the competition between the local shear rate and rotational diffusion (as would occur in a passive rod suspension). This corresponds to the shear-trapping region where cross-streamline migration is very weak due to the strong alignment with the flow. 

However, as we move closer and closer to the centerline, the local shear rate decreases, causing a concomitant decrease in shear alignment and increase in cross-streamline migration due to self-propulsion. This transition corresponds to the edge of the central boundary layer from which particles are depleted, and the position $\delta_{D}$ of this transition region (or half-thickness of the depletion layer) can be estimated by balancing the magnitudes of the terms describing self-propulsion and shear alignment in equation (\ref{eq:goveq1}):
\begin{equation}
\frac{Pe_{s}}{\delta_{D}}\sim \frac{Pe_{f}}{2}\delta_D,
\end{equation}
from which we find
\begin{equation}
\delta_{D}\approx C\sqrt{\chi}, \label{eq:deltaD}
\end{equation}
where the prefactor $C$ is a numerical constant and where we have defined 
\begin{equation}
\chi=\frac{Pe_{s}}{Pe_{f}}=\frac{V_{s}}{2\dot{\gamma}_{w}H}.
\end{equation}
The dimensionless group $\chi$ can be interpreted as the ratio of the timescale $\dot{\gamma}_{w}^{-1}$ it takes a particle to align with the flow  over the characteristic timescale $2H/V_{s}$ it takes it to swim across the channel width: if $\chi$ is small, particles align with the flow much faster than they can cross the channel, leading to significant shear-trapping; on the other hand, if $\chi$ is large, particles cross the channel much faster than they align with the flow and shear-trapping does not occur. As we show in Appendix C, this scaling for $\delta_{D}$ can indeed also be derived by considering the individual trajectories of deterministic swimmers released from the centerline, which can be shown to become trapped at a distance of the order of $\delta_{D}$. It will also be shown to agree quite well with numerical results in \S \ref{sec:trapping}, where we will find that $\delta_{D}\approx  2.404 \sqrt{\chi}$  provides an excellent estimate for the thickness of the depletion layer when $Pe_{s}\lesssim 0.25$ and $Pe_{f}\gtrsim 50$. 

To gain further understanding of the effect of shear rate on the intensity of  depletion, we rescale lengths by $\delta_D$ inside the central boundary layer to rewrite the governing equation (\ref{eq:goveq1}) as
\begin{equation}
\frac{\Gamma}{C}\cos\theta\frac{\partial \Psi}{\partial z}-\textcolor{black}{2\Lambda}\frac{\Gamma^{2}}{C^{2}}\frac{\partial^{2}\Psi}{\partial z^{2}}-\frac{C\Gamma}{\textcolor{black}{2}} z\,  \bnabla_{p}\bcdot\left[\cos\theta (\t{I}-\v{pp})\bcdot\hat{\v{y}}\,\Psi\right] =\frac{1}{2}\nabla_{p}^{2}\Psi, \label{eq:rescaled}
\end{equation}
where the dimensionless group $\Gamma=\sqrt{Pe_{s}Pe_{f}}$ emerges as the most significant parameter governing the profile of the depletion layer. Unsurprisingly, we find that self-propulsion and shear rotation have the same magnitude upon rescaling. In this region, self-propulsion, which scales with $\Gamma$, has the effect of enhancing depletion by driving particles away from the centerline; this competes against translational diffusion, scaling with $\Gamma^{2}$, which has the effect of smoothing concentration gradients and thus hampers depletion. This suggests the following dependence of the concentration profile on $Pe_{f}$. As flow strength is increased from small values, the depletion layer forms and continually narrows according to equation (\ref{eq:deltaD}) for $\delta_D$. As long as $\Gamma<1$, self-propulsion dominates translational diffusion and increasing $Pe_{f}$ (and therefore $\Gamma$) enhances depletion. This trend reverses when $\Gamma\sim O(1)$, when translational diffusion starts to overcome self-propulsion, leading to a subsequent decrease in the strength of depletion for $\Gamma>1$. This qualitative explanation for the non-monotonic dependence of the strength of depletion upon $\Gamma$ (and hence upon the mean shear rate of the imposed Poiseuille flow) is consistent with the experimental observations of \cite{Rusconi14}, and is also borne out by numerical solutions of the governing equations as we describe next.  

\subsection{Arbitrary flow strengths: finite-volume calculations and discussion\label{sec:trapping}}

\begin{figure}
\centering\vspace{0.2cm}
\includegraphics{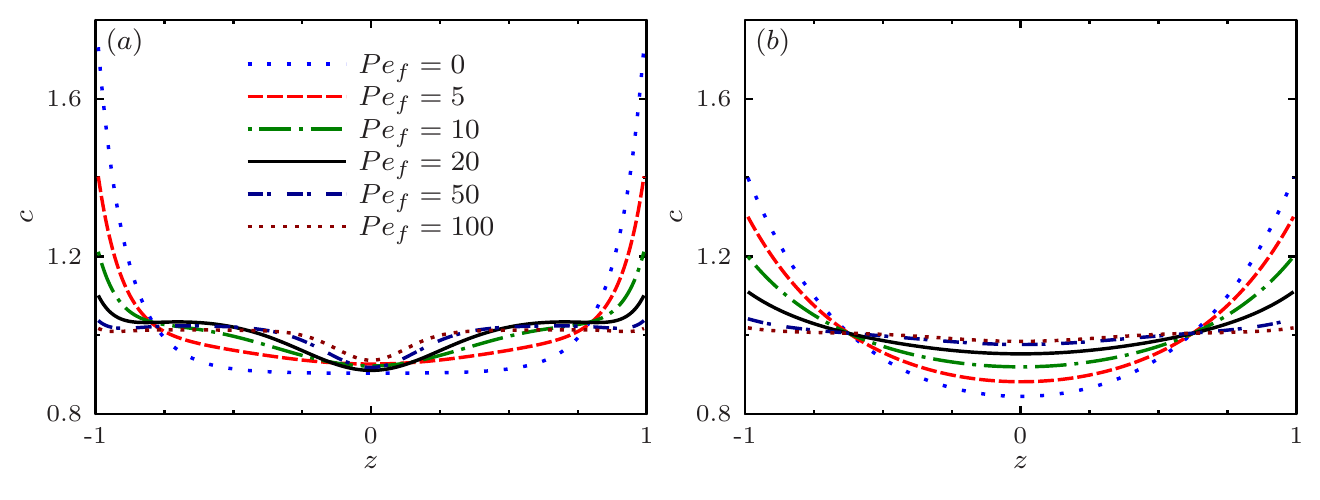}
\caption{(Color online) Equilibrium concentration profiles (at $\Lambda = 1/6$) for (\textit{a}) $Pe_s=0.25$ (strong wall accumulation) and (\textit{b}) $Pe_s=1.0$ (weak accumulation) and for various values of the flow P\'eclet number $Pe_f$, obtained by finite-volume solution of the governing equation (\ref{eq:goveq1}).}\label{fig:arbitraryPec}
\end{figure}

We now test and extend the key predictions from the weak-flow asymptotics and strong-flow scaling analysis from the preceding sections by performing finite-volume numerical simulations of the governing equation (\ref{eq:goveq1}) for arbitrary values of $Pe_{s}$ and $Pe_{f}$ using the algorithm of Appendix C. Typical concentration profiles are illustrated in figure~\ref{fig:arbitraryPec} for various values of $Pe_{f}$, and for the two values of $Pe_{s}=0.25$ and $1.0$ corresponding to cases where wall accumulation in the absence of flow is strong and weak, respectively. In both cases, the leading effect of the external flow on $c$ is to decrease wall accumulation. This trend is easily understood as a result of the alignment of the particles with the flow, which reduces wall-normal polarization and thereby hinders accumulation. This decrease in accumulation also results in a net increase in the concentration in the central parts of the channel and in the flattening of the profiles in the strong-flow limit. When $Pe_{s}$ is small as in figure~\ref{fig:arbitraryPec}(\textit{a}), a depletion layer is also observed to form near the channel centerline and to progressively narrow with increasing $Pe_{f}$, in agreement with the theoretical predictions of \S \ref{sec:strongflow}. At high values of $Pe_{f}$, the three distinct regions identified in \S \ref{sec:strongflow} (wall accumulation, shear-trapping, and centerline depletion) in fact become clearly visible. However, if the swimming P\'eclet number is increased to $Pe_{s}=1.0$ as in figure~\ref{fig:arbitraryPec}(\textit{b}), the thickening of the wall boundary layers suppresses shear-trapping and depletion at the centerline,  leading to a nearly uniform concentration profile in the strong flow limit. 

\begin{figure}
\centering\vspace{0.2cm}
\includegraphics{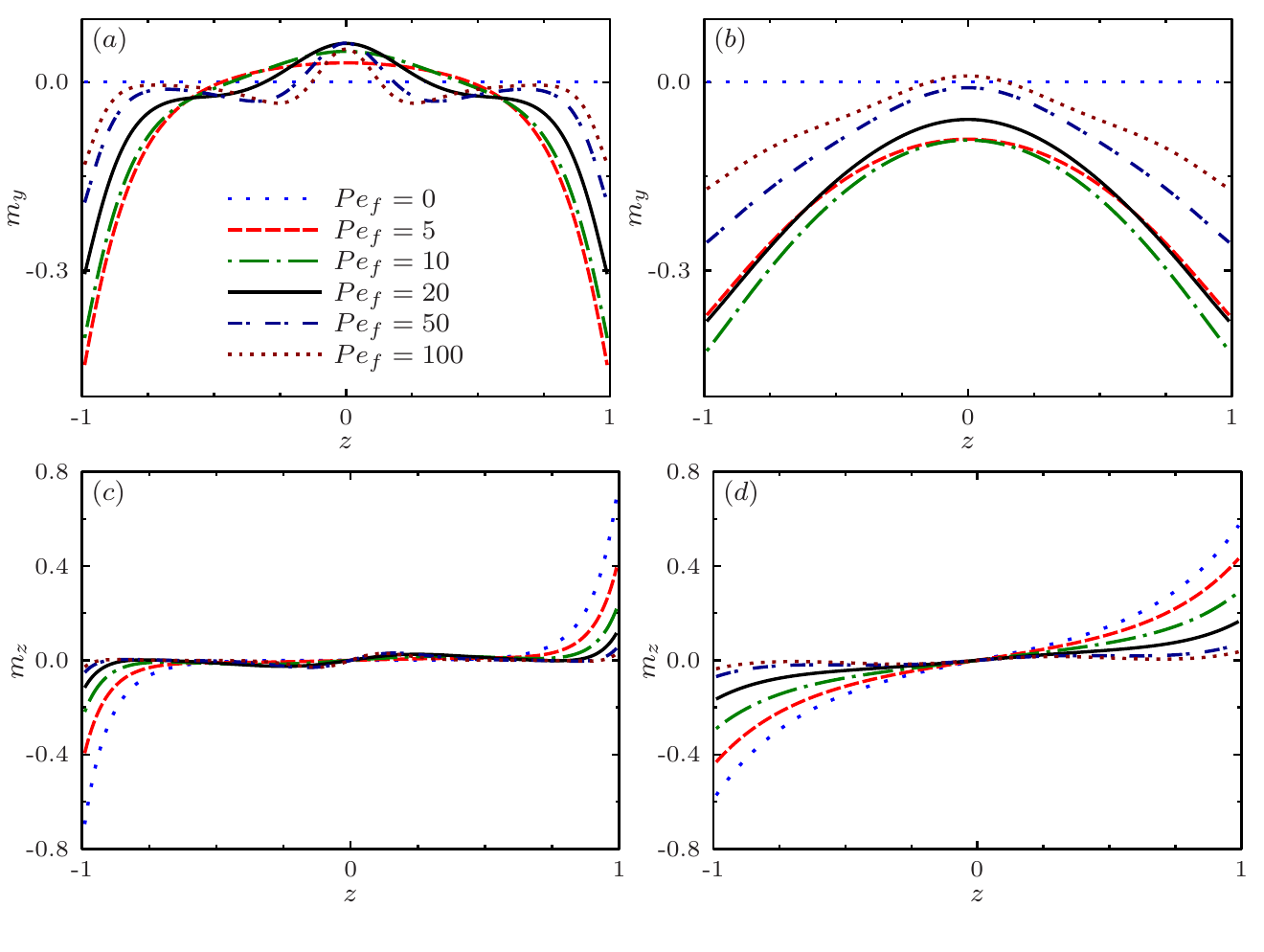}
\caption{(Color online) Equilibrium streamwise and wall-normal polarization profiles (at $\Lambda = 1/6$) for (\textit{a})--(\textit{c}) $Pe_s=0.25$ and (\textit{b})--(\textit{d}) $Pe_s=1.0$ and  for various values of the flow P\'eclet number $Pe_f$, obtained by finite-volume solution of the governing equation (\ref{eq:goveq1}). The streamwise polarization $m_{y}$ is shown on the top row (\textit{a})--(\textit{b}), and the wall-normal polarization $m_{z}$ on the bottom row (\textit{c})--(\textit{d}).}\label{fig:arbitraryPem}
\end{figure}

Corresponding profiles for the wall-normal and streamwise polarization are also shown in figure~\ref{fig:arbitraryPem}. As expected, rotation of the particles by the flow causes a decrease in the wall-normal polarization, and also results in a non-zero streamwise polarization $m_{y}$ as previously discussed in \S \ref{sec:weakflow}. This streamwise polarization is especially strong in the near-wall region where $m_{y}$ is negative, indicating upstream swimming. It is significantly weaker near the center of the channel, where it is found to be positive for $Pe_{s}=0.25$ but remains negative across the entire channel when $Pe_{s}=1.0$ due to the overlap of the two wall boundary layers.

\begin{figure}
\centering\vspace{0.2cm}
\includegraphics{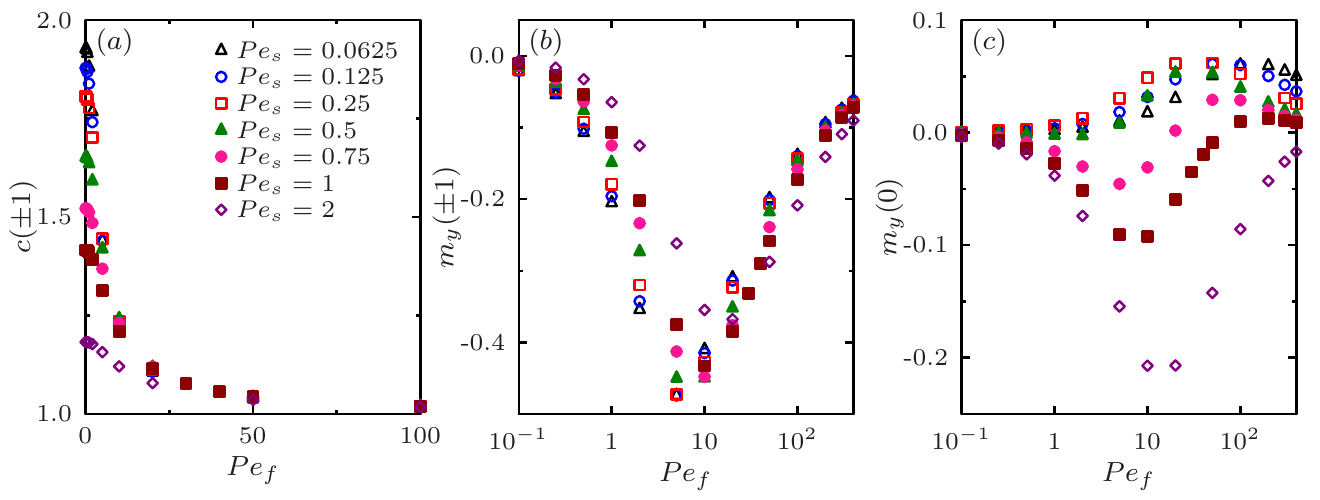}
\caption{(Color online) Effect of swimming and flow P\'eclet numbers on: (\textit{a}) wall concentration $c(\pm1)$, (\textit{b}) streamwise polarization $m_{y}(\pm1)$ at the channel walls, and (\textit{c}) streamwise polarization $m_{y}(0)$ at the channel centerline.}\label{fig:trends}
\end{figure}

These trends are made more quantitative in figure~\ref{fig:trends}, showing the dependence of $c(\pm1)$, $m_{y}(\pm1)$ and $m_{y}(0)$ on the swimming and flow P\'eclet numbers. As previously discussed, the wall concentration is seen to decrease with increasing flow strength irrespective of the value of $Pe_s$, and asymptotically tends to $1$ in the strong-flow limit as the concentration profiles flatten. Figure~\ref{fig:trends}(\textit{b}) shows that the streamwise polarization at the walls is always negative, which implies that the active particles always swim upstream near the boundaries. Interestingly, we find that there is maximum upstream swimming at $Pe_{f}\approx 10$, and the upstream motion is reduced at higher values of the flow P\'eclet number. The streamwise polarization at the channel centerline shows complex trends as shown in figure~\ref{fig:trends}(\textit{c}). As predicted by the weak-flow asymptotic analysis of \S \ref{sec:weakflow}, $m_{y}(0)$ is found to be positive for low values of $Pe_{s}$ and negative for high values of $Pe_{s}$. Its absolute value increases with flow strength in both cases up to $Pe_{f}\approx 10$, beyond which further increasing flow strength reduces the polarization. The decrease in both $m_{y}(\pm1)$ and $m_{y}(0)$ at high $Pe_f$ is a likely consequence of the dominant effect of the shear alignment term in equation (\ref{eq:goveq1}), which promotes nematic rather than polar order. 

\begin{figure}
\centering\vspace{0.2cm}
\includegraphics{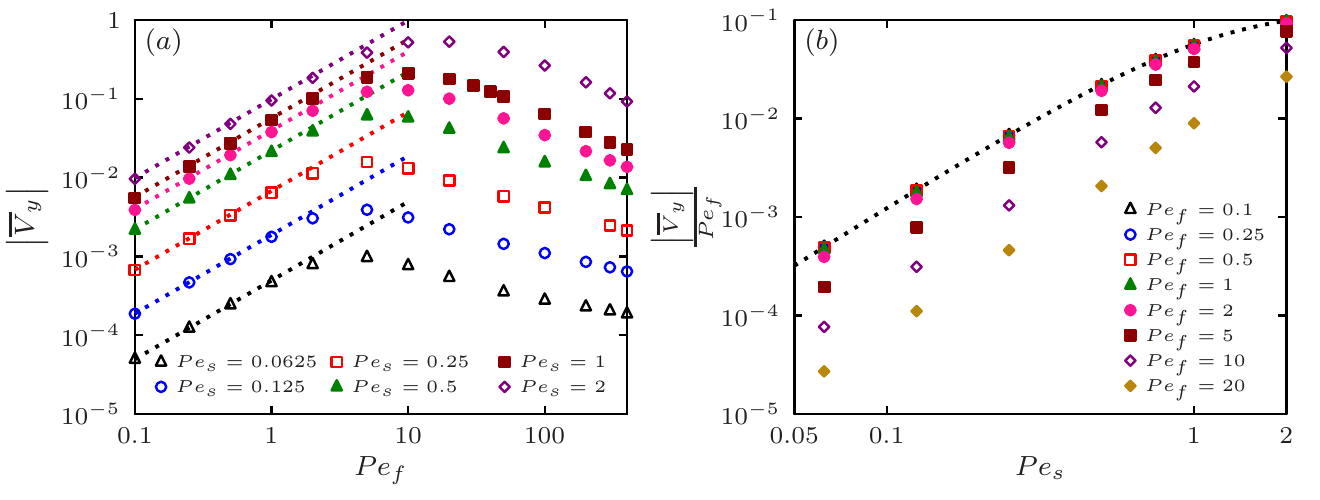}
\caption{(Color online) (\textit{a}) Magnitude of the average upstream swimming velocity $|\overline{V}_{y}|$ as a function of $Pe_f$ for different values of $Pe_{s}$ (at $\Lambda = 1/6$), and (\textit{b}) dependence of $|\overline{V}_y|/Pe_f$ on $Pe_s$ for different values of $Pe_f$. Symbols show finite-volume numerical simulations, and dotted lines show the theoretical prediction of equation (\ref{eq:upvel}).} \label{fig:upswim}
\end{figure}

The dependence of the average streamwise swimming velocity $\overline{V}_y$ defined in equation (\ref{eq:upswim}) on both P\'eclet numbers is shown in figure~\ref{fig:upswim}, where numerical results are  compared to the weak-flow theoretical prediction of equation~(\ref{eq:upvel}). Consistent with figure~\ref{fig:trends}(\textit{b}) for the streamwise polarization at the walls, we find that $\overline{V}_{y}<0$, and that $|\overline{V}_{y}|$ first increases nearly linearly with $Pe_{f}$  in agreement with the predictions of \S \ref{sec:weakflow}. This increase persists up to $Pe_f\approx 10$, beyond which $|\overline{V}_{y}|$ starts decreasing again. Excellent quantitative agreement is found with equation (\ref{eq:upvel}) for $Pe_f\lesssim 2.0$. This is confirmed in figure~\ref{fig:upswim}(\textit{b}), showing the dependence of $|\overline{V}_y|/Pe_{f}$ on swimming P\'eclet number: the upstream velocity is found to increase with $Pe_{s}$, primarily as a result of the corresponding increase in swimming speed of individual particles, and a collapse of all the curves onto the theoretical prediction of equation (\ref{eq:upvel})  is observed when $Pe_f\lesssim 2.0$.

\begin{figure}
\centering\vspace{0.2cm}
\includegraphics{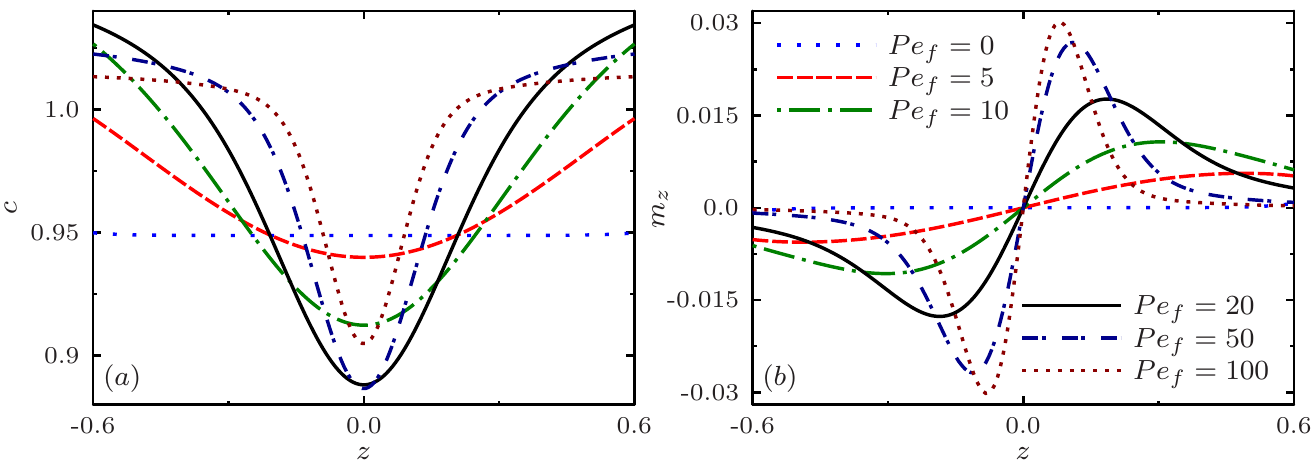}
\caption{(Color online) (\textit{a}) Concentration profiles in the central portion of the channel for $Pe_{s}=0.125$ and various values of the flow P\'eclet number $Pe_{f}$, obtained by finite-volume solution of equation (\ref{eq:goveq1}). (\textit{b}) Corresponding profiles of the wall-normal polarization $m_{z}$.  } \label{fig:trapping}
\end{figure}

As seen in figure~\ref{fig:arbitraryPec}(\textit{a}), shear-trapping and centerline depletion are observed in the central portion of the channel at high flow P\'eclet number if $Pe_s$ is sufficiently low. This is illustrated more clearly in figure~\ref{fig:trapping}, where concentration and wall-normal polarization profiles are shown in the central portion of the channel for various values of the flow P\'eclet number and for $Pe_s=0.125$. This value was chosen to match the experiments of \cite{Rusconi14}, where the following parameters were reported: $V_s=50\,\mu$m, $d_r=1$~s$^{-1}$, and $2H=400\,\mu$m. As seen in figure~\ref{fig:trapping}(\textit{a}), increasing $Pe_{f}$ from zero first results in a decrease in the concentration at the centerline, corresponding to the formation of the depletion layer. As the concentration decreases, the width of the depletion layer is also found to decrease. This trend continues up to $Pe_{f}\approx 20$, above which the concentration at the centerline starts increasing again, even though the depletion layer keeps narrowing. These trends are in very good agreement with the experiments of \cite{Rusconi14}, who also reported a non-monotonic dependence of the strength of depletion on shear rate; in fact, the profiles shown in figure~(\ref{fig:trapping}) are very similar to the experimental profiles at equivalent values of $Pe_{f}$. The trends on the concentration profile are easily understood based on figure~\ref{fig:trapping}(\textit{b}) for the wall-normal polarization, which reflects the net swimming velocity across the channel and provides insight into cross-streamline migration.  Indeed, the polarization profiles exhibit peaks on both sides of the depletion layer, corresponding to a strong migration away from the center. These peaks increase in magnitude and also shift towards the centerline as flow strength increases and the depletion layer narrows. Beyond those peaks, $m_{z}$ quickly decays to zero where the concentration profiles plateau in accordance with equation (\ref{eq:cmzrel}) and shear-trapping of the particles takes place. 

\begin{figure}
\centering\vspace{0.cm}
\includegraphics{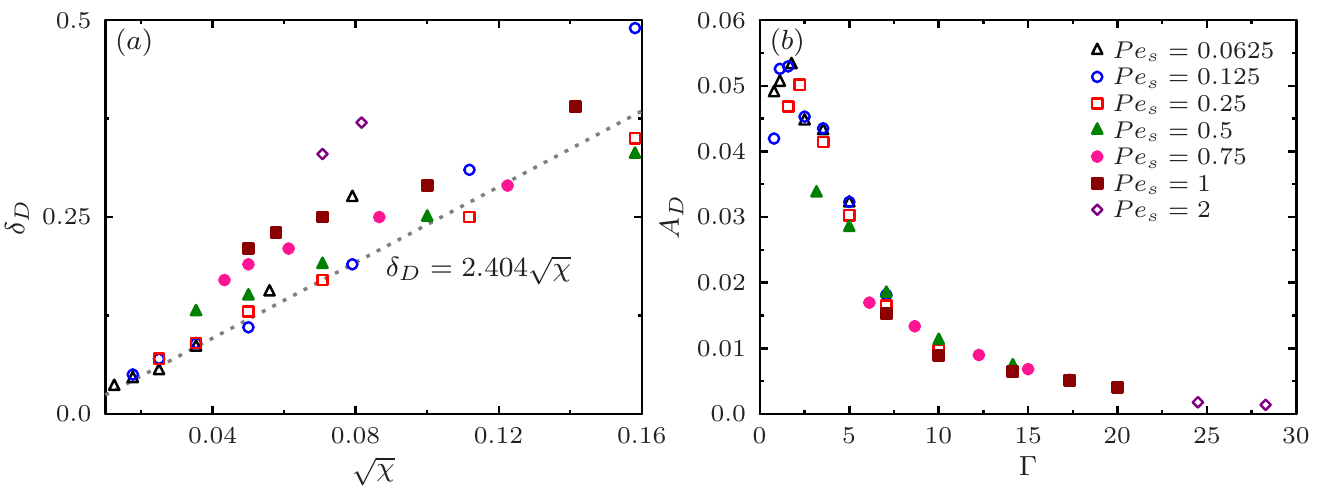}
\caption{(Color online) (\textit{a}) Depletion layer thickness $\delta_D$, defined as the distance from the centerline where the wall-normal polarization reaches its maximum, as a function of $\sqrt{\chi}=\sqrt{Pe_s/Pe_f}$. (\textit{b}) Depletion index $A_D$ defined in equation (\ref{eq:index}) as a function of $\Gamma=\sqrt{Pe_sPe_f}$. 
} \label{fig:trapping2}
\end{figure}

These trends are tested more quantitatively against the strong-flow scaling analysis of \S \ref{sec:strongflow} in figure~\ref{fig:trapping2}. We first define the thickness $\delta_{D}$ of the depletion layer as the distance from the centerline where $m_{z}$ reaches its maximum, when such a maximum exists. Based on the analysis of \S \ref{sec:strongflow}, we expect $\delta_{D}$ to scale linearly with $\sqrt{\chi}=\sqrt{Pe_s/Pe_f}$ in strong flows, and this is indeed confirmed in figure~\ref{fig:trapping2}(\textit{a}). We find that $\delta_{D}$ can only be defined when $\sqrt{\chi}\lesssim 0.16$ or $Pe_f\gtrsim 40Pe_{s}$, which corresponds to the shear-trapping regime. Best agreement with the scaling prediction is obtained in the low $Pe_{s}$ and high $Pe_f$ limit, and a linear least-square fit to the data for $Pe_{s}\le 0.25$ and $Pe_f\ge 50$ shows that $\delta_{D}\approx 2.404\sqrt{\chi}$. As $Pe_s$ increases, the numerical results depart from this prediction, primarily due to the thickening of the wall boundary layers which causes them to interact with the parts of the channel where shear-trapping and depletion occur. We further quantify the shape of the depletion layer by introducing a depletion index $A_D$ measuring the amount of particles depleted from the center due to trapping in high-shear regions:
\begin{equation}
A_D=\int_{0}^{\delta_{D}}c(z)\,\mathrm{d}z-\delta_{D}c(\delta_{D}). \label{eq:index}
\end{equation}
As we argued in \S \ref{sec:strongflow} based on equation (\ref{eq:rescaled}), the shape of the depletion layer is expected to depend upon $\Gamma=\sqrt{Pe_s Pe_f}$, and indeed the numerical data for the depletion index for various values of $Pe_{s}$ and $Pe_{f}$ is found to collapse onto a master curve when plotted vs $\Gamma$ in figure~\ref{fig:trapping2}(\textit{b}).  In agreement with the trends observed in figure~\ref{fig:trapping}(\textit{a}), the depletion index shows a non-monotonic dependence on $\Gamma$, with maximum depletion occurring for $\Gamma\approx 2$. 

\begin{figure}
\centering\vspace{0.cm}
\includegraphics[width=9.8cm]{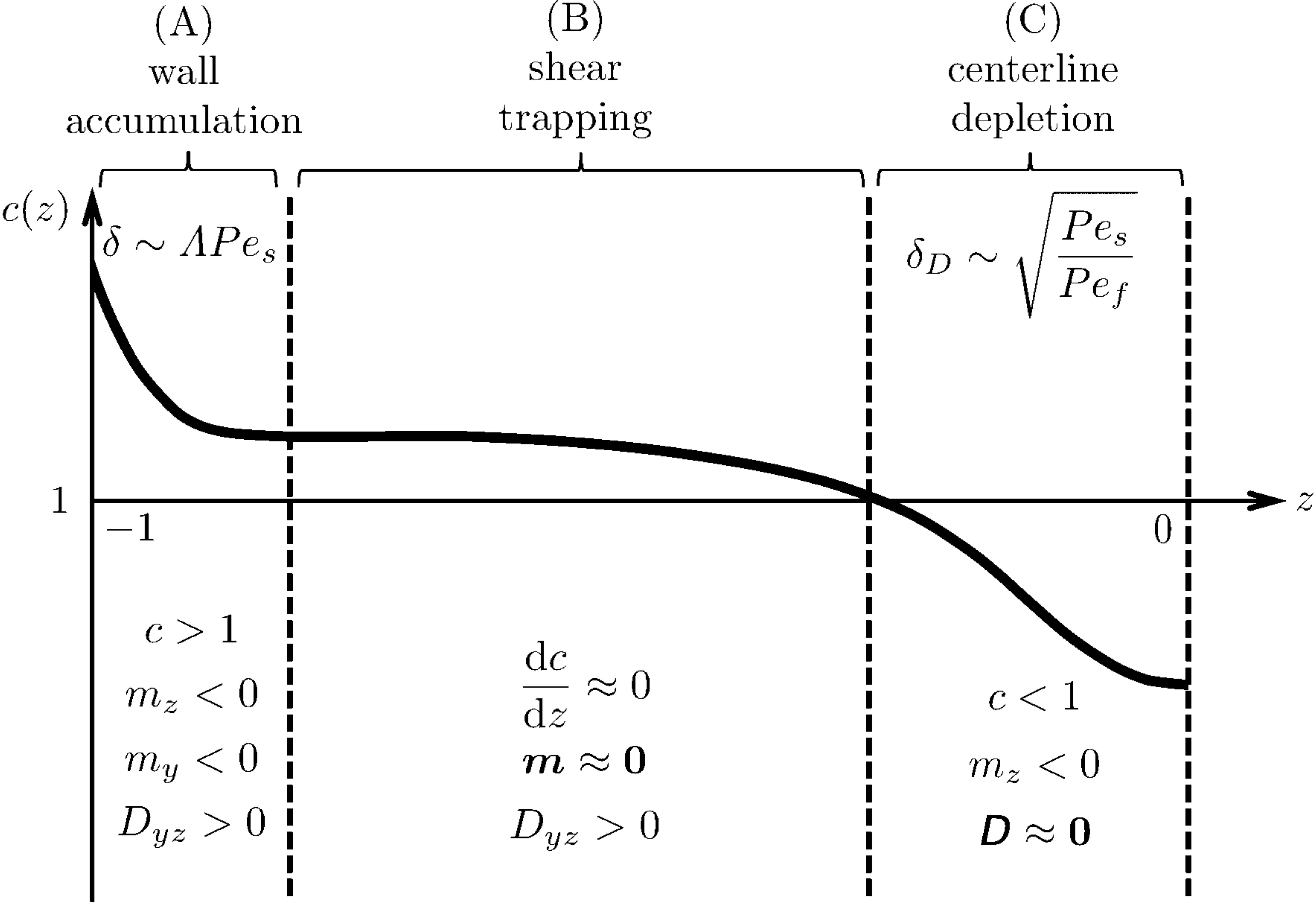}
\caption{Schematic summary of the dynamics in the limits of $Pe_{s}\ll 1$ and $Pe_{f}\gg 1$.  The channel can be roughly divided into three regions: (A) near the walls, particles accumulate in a boundary layer of thickness $\delta\sim \Lambda Pe_{s}$; (B) away from the walls and centerline, strong nematic alignment by the flow leads to shear-trapping and a nearly uniform concentration profile; (C) near the centerline, particle propulsion leads to a depletion layer of thickness $\delta_{D}\sim \Gamma$. The diagram only shows the left half of the channel $z\in[-1,0]$; the corresponding diagram in the other half can be obtained by symmetry and by noting that $m_{z}$ is an even function of $z$, whereas $m_{y}$ and $D_{yz}$ are both odd functions.} \label{fig:trapping3}
\end{figure}

The dynamics in the limits of $Pe_{s}\ll 1$ and $Pe_{f}\gg 1$ are summarized schematically in figure~\ref{fig:trapping3}, where the channel can be roughly divided into three distinct regions. Region (A), with thickness  $\delta\sim \Lambda Pe_{s}$, abuts the channel wall and is characterized by wall accumulation and a net polarization towards the wall. These effects occur even in the absence of flow, and are in fact mitigated by the flow which tends to decrease the wall concentration and rotate particles to induce upstream polarization. Away from both the wall and the channel centerline is region (B), where the concentration profile is nearly uniform and shear trapping occurs: here, polarization is weak but there is a strong nematic alignment of the particles due to the applied shear. The local shear rate decreases in magnitude as we approach the centerline and enter region (C), which has a characteristic thickness of $\delta_{D}\sim\sqrt{Pe_{s}/Pe_f}$: in this region, particles are depleted due to a net polarization towards the walls, which drives migration away from the center but is counterbalanced by translational diffusion. Increasing $Pe_{s}$ causes both regions (A) and (C) to widen, up to a point where they merge and the three regions can no longer be distinguished. Increasing $Pe_f$, on the other hand, tends to weaken wall accumulation but does not change the thickness of region (A), while it also causes the narrowing of region (C).

\section{Discussion\label{sec:conclusions}}

\subsection{Summary of main results}

We have used a combination of theory and numerical simulations to analyze the distributions and transport properties  of an infinitely dilute suspension of self-propelled particles confined between two parallel flat plates, both in quiescent conditions and under an imposed pressure-driven flow. Our analysis focused on incorporating the effects of confinement within the kinetic theory framework previously developed by \cite{SS2008a}, which is based on a Smoluchowski equation for the distribution of the active particle positions and orientations. In particular, we demonstrated that prescribing a zero-normal-flux condition on the  particle distribution function at the boundaries captures several key features reported in experiments on dilute active suspensions under confinement. We presented a finite-volume algorithm for the numerical solution of the Smoluchowski equation, which allows for an easy implementation of the boundary conditions, and also developed a simpler system of equations for the orientational moments of the distribution function, which enabled us to perform analytical calculations in the absence of flow and under a weak imposed flow. An asymptotic scaling analysis was also performed on the full Smoluchowski equation under strong flow. The numerical simulation data was used to test and further understand the analytical calculations and predictions.  

We first considered the dynamics in the absence of flow. In this case, the governing equations involve a  swimming P\'eclet number $Pe_{s}$, which is the ratio of the persistence length of swimmer trajectories to the channel height, \textcolor{black}{as well as a parameter $\Lambda$ that is fixed for a given swimmer type and whose inverse measures the strength of propulsion}.  In the \textcolor{black}{limit of wide channels}, the channel can be divided into two regions: a near-wall accumulation region where the particles tend to concentrate and have a net polarization towards the wall, and a bulk region away from the walls where the distribution is nearly uniform and isotropic. \textcolor{black}{Asymptotic expressions for the full distribution function were also derived as  series in powers of $\Lambda$ in the weak and strong propulsion limits. In particular, it was shown that the characteristic thickness of the accumulation layer scales with $d_{t}/V_{s}$ in the strong propulsion limit ($\Lambda \ll 1$), and with $\sqrt{d_{t}/d_{r}}$ in the weak propulsion limit ($\Lambda\gg 1$). For finite values of $\Lambda$,}  analytical expressions for the concentration and polarization profiles were obtained by solving the moment equations and displayed excellent agreement with the finite-volume numerical simulation of the full distribution function for a wide range of  values of the swimming P\'eclet number \textcolor{black}{so long as $\Lambda \gtrsim 0.1$. }
Based on these results, we proposed and validated a simple mechanism for wall accumulation, where the presence of the wall breaks the polar symmetry of the active particles and leads to sorting of orientations. This mechanism differs from previous explanations based on hydrodynamic interactions or surface alignment due to collisions, and led us to conclude that both pusher and puller particle suspensions will exhibit similar wall accumulation in the dilute limit. Hydrodynamic and surface alignment interactions are, however, expected to quantitatively affect the profiles \textcolor{black}{in more concentrated systems} and to lead to different distributions for pusher and puller particles.

Next, we analyzed the effects of an imposed pressure-driven flow. When a flow is applied on the suspension, the physics is now governed by \textcolor{black}{three} dimensionless groups: the swimming P\'eclet number $Pe_s$ \textcolor{black}{and parameter $\Lambda$} introduced above, as well as a flow P\'eclet number $Pe_{f}$ comparing the imposed shear rate to rotational diffusion. In the weak flow limit, we calculated the leading-order corrections of the streamwise polarization and shear nematic alignment due to the flow and showed that near-wall upstream swimming is a consequence of shear rotation of the particles inside the accumulation layer near the walls. We  derived an analytical expression for the average upstream swimming velocity of the active particles relative to the imposed flow, which was compared against numerical simulations and provides an excellent estimate for $Pe_{f} \lesssim 2$. In the strong flow limit, we developed a scaling analysis to show that when  $Pe_{s}\ll 1$ and $Pe_{f}\gg 1$ the channel can be roughly divided into three regions: the near-wall accumulation region with thickness $\delta \sim \Lambda Pe_{s}$, a depletion region near the centerline with thickness $\delta_{D}\sim\Gamma=\sqrt{Pe_{s}/Pe_f}$, and a shear-trapping region away from the wall and centerline where the concentration is nearly uniform and particle alignment is primarily nematic. The extent of the central depletion shows a non-monotonic variation with flow strength, with a maximum depletion occurring at a critical flow strength such that $\Gamma \sim O(1)$.

\subsection{\textcolor{black}{Discussion and comparison to previous works}} 

\textcolor{black}{The phenomena analyzed in this study have received considerable attention in experiments as well as other models and simulations, so we compare and contrast them here to these prior works. As mentioned in the introduction, the wall accumulation predicted by our model in the absence of flow is well known in experiments on bacterial suspensions, where accumulation layers of $\approx 1$ to 50 $\mu$m are typically reported \citep{Berke08,Li09,Li11,Gachelin14}, with increases in concentration of up to 50 times the bulk density very close to the wall \citep{Li11}. Such high concentrations at the walls are consistent with our numerical results of figure~\ref{fig:noflow_Lambda}, which predict high values of $c(\pm1)$ in the strong-propulsion limit of $\Lambda\ll 1$ relevant to bacteria. Indeed, a rough estimate for \textit{E. coli} provides $\Lambda\approx 0.01$, though it is difficult to precisely measure $d_{t}$ in experiments since long-time mean-square displacements are dominated by Taylor dispersion. This strong accumulation is also consistently observed in simulations \citep{Hernandez05,Nash10,Costanzo12,Elgeti13,Lushi14,Li14}, which also exhibit the preferential alignment of the swimmers towards the wall that our model predicts. A similar alignment has also been reported in a few experiments \citep{Drescher11,Lushi14}, though detailed observations of  swimming micro-organisms near walls has also revealed complex complex scattering dynamics due to the interactions of the flagellar appendages with the boundaries \citep{Denissenko12,Kantsler13}. These observations seem to contradict mechanisms purely based on Stokes-dipole hydrodynamic interactions with the no-slip walls, as these predict reorientation of the cells parallel to the walls in the case of pushers \citep{Berke08}. Rather, they appear to support the prediction that accumulation layers derive predominantly from a polarity-sorting mechanism across the channel together with a balance of self-propulsion and diffusion at the walls. We note that this mechanism was also proposed in the work of \cite{Elgeti13}, who performed simulations of self-propelled Brownian spheres between two flat plates. Their numerical results support the trends described in \S \ref{sec:numres} on the effect of confinement as captured by $Pe_{s}$. \cite{Elgeti13} also wrote down a continuum model that shares similarities with ours, which they used to analyze the strong propulsion and narrow gap limits. Their conclusions are in agreement with the discussion of \S \ref{sec:small_L} and~\S \ref{sec:large_L}.}

\textcolor{black}{The distributions and dynamics predicted by our theory under imposed flow also agree with the bulk of prior studies, both experimental and numerical. The reorientation of near-wall swimmers against the flow leading to upstream swimming has been reported ubiquitously in many experiments \citep{Hill07,Kaya09,Kaya12,Kantsler14}  and simulations \citep{Nash10,Costanzo12,Chilukuri14}, with several of these studies proposing similar mechanisms as that described herein, namely the shear rotation of the polarized cells near the walls. Quite remarkably, the peak in the upstream swimming flux at a critical flow strength visible in the simulation data of figure~\ref{fig:upswim}(\textit{a}) was also reported in the experiments of \cite{Kantsler14}.}

\textcolor{black}{The dynamics in strong flows in the central part of the channel has only received little attention in previous studies. Our interest in this problem was sparked by the recent microfluidic experiments of \cite{Rusconi14}, which were the first to predict centerline depletion and shear trapping. Our scaling analysis and numerical results of \S \ref{sec:strongflow} and \S \ref{sec:trapping} are in excellent agreement with their observations. In particular, the shape of the concentration profiles near the channel centerline obtained in figure~\ref{fig:trapping} are quite similar to those shown in figure 2(\textit{a}) of their paper. Further, we observed in our study a non-monotonic dependence of the depletion index on $\Gamma$, with maximum depletion occurring for $\Gamma\approx 2$. In the experiments of \cite{Rusconi14}, a similar non-monotonic trend was reported, with the strongest depletion occurring in the range of $\gamma\approx 2.5$ -- $10$ s$^{-1}$. From their data, we estimate $Pe_{f}\approx 5$ -- $20$ and $Pe_{s}\approx 0.125$, from which we find $\Gamma\approx 0.8$ -- $1.6$ in reasonable agreement with our numerical results. A simple analytical model based on a Fokker-Planck equation was also introduced in their paper, though only limited results were obtained in the low-$Pe_{f}$ limit. }

\textcolor{black}{Since the experiments of \cite{Rusconi14}, the existence of centerline depletion in strong flows was also confirmed in the numerical simulations of \cite{Chilukuri14}, which provided additional insight into the shape of the depletion layer and its scaling with flow strength. By fitting the dip in concentration at the centerline with a parabola, they were able to extract the profile curvature from their simulation data, and showed that it collapses onto a master curve when plotted vs $\dot{\gamma}_{w} H/2V_{s}$, in agreement with our prediction that the shape of the depletion is controlled by $\chi=Pe_{s}/Pe_{f}=V_{s}/2\dot{\gamma}_{w}H$. Their also reported similar particle orientations as predicted in figures~\ref{fig:weakflow}(\textit{a}) and \ref{fig:arbitraryPem}(\textit{a}): namely, swimmers are preferentially aligned with the flow in the bulk of the channel, even though they tend to swim upstream near the walls. Finally, we recall that our theoretical scaling for the width of the depletion layer is also in agreement with the analytical model of \cite{Zottl12}, which is discussed in more detail in Appendix D and determines the distance away from the centerline where a deterministic swimmer leaving $z=0$ with a given orientation fully aligns with the flow, i.e., becomes trapped by shear alignment.}

\subsection{Concluding remarks}

The favorable agreement of our predictions with both experiments and simulations validates our model and in particular our choice of boundary condition. We reiterate that particle-particle and particle-wall hydrodynamic interactions were entirely neglected in this work, suggesting that the salient features of confined active suspensions such as wall accumulation, upstream swimming, centerline depletion and shear-trapping can all be explained in the absence of such interactions. Yet even in dilute suspensions, particle-wall hydrodynamic interactions are known play a role  \citep{Spagnolie12} and are expected to slightly modify the results described here. Pusher and puller suspensions are no longer equivalent when hydrodynamic interactions are included and therefore may adopt slightly different distributions, whereas this distinction is irrelevant in the present model. As particle density increases, we also expect particle-particle hydrodynamic interactions to become significant, and to destabilize the equilibrium distributions obtained in \S \ref{sec:noflow} if the concentration is sufficiently high. A preliminary one-dimensional stability analysis accounting for flow modification by the particles suggests the existence of a symmetry-breaking bifurcation above a critical concentration in suspensions of pushers, leading to unidirectional flow with net fluid pumping; such an instability was also previously predicted using various phenemenological models for active liquid crystals \textcolor{black}{\citep{Voituriez05,Edwards09,Ravnik13, Furthauer12,Marenduzzo07a}}. Further increases in concentration may also lead to the onset of bacterial turbulence \textcolor{black}{ \citep{Marenduzzo07b,Gachelin14}. These predictions have yet to be confirmed from a hydrodynamics first-principles perspective and may also be investigated computationally using a generalization of the finite-volume algorithm presented in Appendix~C, or by numerical solution of the approximate equations for the orientational moments of the distribution function, which were shown to be highly accurate in the absence of an external flow. Since the equilibrium states under confinement are non-uniform and polarized in the wall-normal direction, the instabilities in confined active suspensions could have multifold origins.} 

\textcolor{black}{Our study has only focused on the limit of high-aspect-ratio particles whose orientational dynamics are described by equation~(\ref{eq:pdot}). If the aspect ratio of the particles is not high, some of the conclusions of this work may change.  The distributions in the absence of flow, including the formation and structure of the wall accumulation layers, are not expected to change even in the limit of spherical particles, as confirmed by previous simulations of Brownian active spheres \citep{Elgeti13}. However, small-aspect-ratio particles will be subject to a weaker alignment with the local shear in an imposed flow, which is expected to widen and eventually suppress the centerline depletion layer in strong flows. This concept may provide interesting avenues for the sorting of active particles by shape in microfluidic devices.}

\textcolor{black}{As a final comment, we recall that a crucial ingredient of our analysis is the presence of translational diffusion in the dynamics of the swimmers, which acts to balance the swimming flux at the boundaries and leads to diffuse accumulation layers. In the limit of strong propulsion or weak diffusion ($\Lambda\rightarrow 0$), we saw that accumulation is enhanced, and we expect the formation of concentration singularities at the walls in the strict limit of $d_{t}=0$. This limit is not easily addressed in the context of our theory, though a very recent attempt at describing accumulation in this case was proposed by \cite{Elgeti15}. The development of a more detailed framework in the absence of diffusion may prove particularly relevant for describing the accumulation of fast-swimming bacteria undergoing run-and-tumble dynamics, notably in applications involving the interaction of bacterial suspensions with suspended passive objects \citep{Sokolov10,Dileonardo10,Koumakis13,Kaiser14}. }

\begin{acknowledgments}
The authors thank \textcolor{black}{John Brady}, Anke Lindner, Eric Cl\'ement, Roman Stocker, \textcolor{black}{Roberto Rusconi}  and Jeffrey Guasto for useful conversations on this problem. D.S. gratefully acknowledges funding from NSF CAREER Grant No.\ CBET-1151590.
\end{acknowledgments}

\appendix
\vspace{0.5cm}

\section{\textcolor{black}{Comparison between the no-flux and reflection boundary conditions}}

\textcolor{black}{In this Appendix, we compare the no-flux boundary condition of equation~(\ref{eq:BC0}), which is central to our model, to the reflection boundary condition used in previous works \citep{Ezhilan12, Bearon11}. The reflection boundary condition ensures	 that 
\begin{equation}
\Psi\left(\pm 1, \theta, \phi\right) = \Psi\left(\pm 1, \pi - \theta, \phi\right), \label{eq:reflBC}
\end{equation}
at the channel walls, where $\theta$ and $\phi$ are defined in Figure~1.  Calculating the first three orientational moments of equation (\ref{eq:reflBC}) yields the following conditions to be enforced at $ z = \pm 1$:
\begin{align}
&\frac{\mathrm{d}c}{\mathrm{d}z}= 0, \label{eq:c2BC}\\
& m_{z} = 0, \quad \frac{\mathrm{d}m_{y}}{\mathrm{d}z}= 0, \\
&\frac{\mathrm{d}D_{zz}}{\mathrm{d}z}= 0, \quad \frac{\mathrm{d}D_{yy}}{\mathrm{d}z}=0, \quad D_{yz} = 0. \label{eq:D2BC}
\end{align}
While equations (\ref{eq:c2BC})--(\ref{eq:D2BC}) are easily shown to imply that the no-flux conditions (\ref{eq:cBC})--(\ref{eq:DBC}) on $c$, $m_{y}$,$D_{yy}$, $D_{zz}$ are also satisfied, they are much more stringent conditions,  with a significant impact on the distribution of particles near the wall.}

\textcolor{black}{First, in the absence of flow, we see that equations~(\ref{eq:ceq1})--(\ref{eq:Dzzeq1}) now need to be solved subject to boundary conditions (\ref{eq:c2BC})--(\ref{eq:D2BC}) at $z = \pm 1$. The uniform and isotropic solution with $c^{(0)}=1$ and  $m_{z}^{(0)}=D_{zz}^{(0)}=0$ satisfies this system exactly. In other words, the condition of \ref{eq:reflBC}, by enforcing a zero concentration gradient and wall-normal polarization at the walls, is unable to capture the concentration/polarization boundary layer which is one of the key results predicted by the no-flux boundary condition and is a ubiquitous feature of experiments and particle models.}

\textcolor{black}{The impact of condition (\ref{eq:reflBC}) on distributions under flow can be understood in the low $Pe_{f}$ limit by modifying the derivation of \S\ref{sec:weakflow}. Since $m_{z}^{(0)}=0$, the right-hand term in equation~(\ref{eq:my1eq}) now vanishes. Equation~(\ref{eq:my1eq})--(\ref{eq:Dyz1eq}) are then rewritten as
\begin{align}
Pe_{s}\frac{\mathrm{d}D_{yz}^{(1)}}{\mathrm{d}z}-2\Lambda{Pe}_{s}^{2}\frac{\mathrm{d}^{2}m_{y}^{(1)}}{\mathrm{d}z^{2}}+m_{y}^{(1)}&=0, \label{eq:my1eqD}\\
\frac{Pe_{s}}{5}\frac{\mathrm{d}m_{y}^{(1)}}{\mathrm{d}z}-2\Lambda Pe_{s}^{2}\frac{\mathrm{d}^{2}D_{yz}^{(1)}}{\mathrm{d}z^{2}}+3D_{yz}^{(1)}&= \frac{S(z)}{10}, \label{eq:Dyz1eqD}
\end{align}
subject to the boundary conditions
\begin{equation}
\frac{\mathrm{d}m_{y}^{(1)}}{\mathrm{d}z}=0, \quad D_{yz}^{(1)}=0 \quad \mathrm{at}\,\,\, z=\pm1. \label{eq:BCmy1D}
\end{equation}
Taking a cross-sectional average of equation~(\ref{eq:my1eqD}) subject to equation~(\ref{eq:BCmy1D}) shows  that $\overline{m}^{(1)}_{y} = 0$. Therefore, the mean upstream velocity in the channel is exactly zero if the reflection boundary condition is enforced. The condition also imposes a zero streamwise nematic alignment ($D_{yz}^{(1)}=0$) at the walls, which is not physical when a fluid flow satisfying the no-slip boundary condition is imposed. A closer look at equations~(\ref{eq:Dyz1eqD})--(\ref{eq:BCmy1D}) also reveals that the system is in fact ill-posed in the limit of $Pe_{s} \to 0$. For finite values of $Pe_{s}$, a numerical solution shows that the reflection boundary condition severely underpredicts the near-wall upstream polarization shown in figure~(\ref{fig:weakflow}). Finally, we note that the analysis presented in \S\ref{sec:strongflow} in the strong-flow limit (and hence the scalings for the depletion boundary layer thickness and rationalization of the non-monoticity of the depletion index with $Pe_{f}$) describe the dynamics in the bulk of the channel and is not affected by the boundary condition imposed.}

\section{Effect of steric exclusion}

The analysis of this paper entirely neglected the finite size of the active particles and in particular did not account for steric exclusion with the boundaries, which is expected to modify the distributions near the walls as observed experimentally \citep{Takagi14}. As previously shown in the case of passive rods \citep{Nitsche90,Schiek95,Krochak10}, excluded volume interactions can be incorporated by means of a more complex boundary condition. One must first realize that steric exclusion prohibits those configurations near either of the two walls that lead to overlap of a section of a particle with the wall. The boundaries between such allowed and prohibited configurations define two hypersurfaces in the three-dimensional $(z,\theta,\phi)$ space of particle configurations:
\begin{align}
& z=1-L^{*}\,|\hspace{-0.05cm}\cos\theta| \,\,\,\,\qquad \mathrm{(top\,\, hypersurface),}  \label{eq:hyper1}   \\
& z=-1+L^{*}\,|\hspace{-0.05cm}\cos\theta|\qquad \mathrm{(bottom\,\, hypersurface),}   \label{eq:hyper2}
\end{align}
where $L^{*}=L/2H$ is the ratio of the particle length to the channel width. At any position $z$ inside the channel, this restricts the allowable range of $\theta$ to an interval of the form $[\theta_{1}(z),\theta_{2}(z)]$, with
\begin{equation}
\theta_{1}(z) = \begin{cases}  \,0 &\mbox{ for }  1-|z| \ge L^{*}, \\  \,\cos^{-1}\left(\displaystyle\frac{1-|z|}{L^{*}}\right) &\mbox{ for } 1-|z| \le L^{*}, \end{cases}
\end{equation}
and
\begin{equation}
\theta_{2}(z) = \begin{cases}  \,\upi &\mbox{ for }  1-|z| \ge L^{*}, \\  \,\cos^{-1}\left(\displaystyle\frac{-1+|z|}{L^{*}}\right) &\mbox{ for } 1-|z| \le L^{*}, \end{cases}
\end{equation}
and consequently, any integral with respect to $\v{p}$ of a field variable $A(z,\v{p})$ must be restricted to these configurations:
\begin{equation}
\int_{\Omega}A(z,\v{p})\,\mathrm{d}\v{p}\equiv \int_{0}^{2\upi}\int_{\theta_{1}(z)}^{\theta_{2}(z)}A(z,\v{p})\sin\theta\,\mathrm{d}\theta\,\mathrm{d}\phi. 
\end{equation}

\begin{figure}
\centering\vspace{0.0cm}
\includegraphics[scale=1.2]{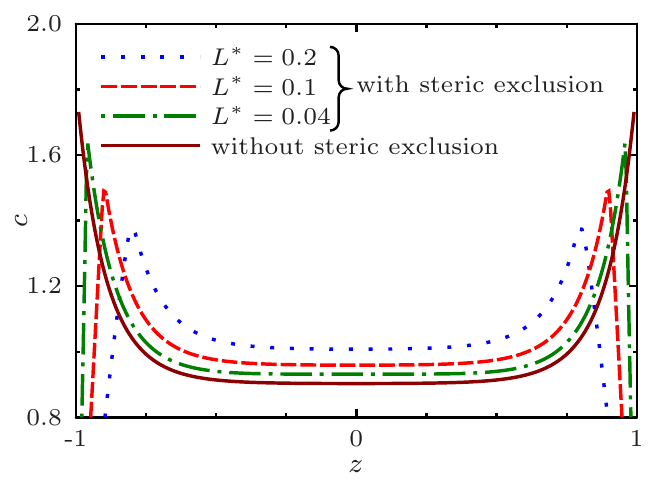}
\caption{(Color online) Effect of steric exclusion on the steady concentration profile in the absence of flow and for $Pe_{s}=0.25$. The plot compares numerical results for three different values of $L^{*}=L/2H$ to the case where steric exclusion is neglected ($L^{*}\rightarrow 0$).  } \label{fig:exclusion}
\end{figure}

To ensure that prohibited configurations are never realized, the boundary condition (\ref{eq:BC00}) must be replaced by a more general no-flux condition on the hypersurfaces defined in equations (\ref{eq:hyper1})--(\ref{eq:hyper2}). Introduce the generalized flux vector $\v{J}$ as
\begin{equation}
\v{J}(z,\v{p},\Psi)=(\dot{\v{x}}+\dot{\v{p}})\,\Psi=J_{z}\hat{\v{z}}+J_{\theta}\hat{\boldsymbol{\theta}}+J_{\phi}\hat{\boldsymbol{\phi}}, \label{eq:J0}
\end{equation}
with 
\begin{align}
J_{z}&=Pe_{s}\cos\theta \,\Psi-\textcolor{black}{2\Lambda} Pe_{s}^{2}\frac{\partial \Psi}{\partial z}, \\
J_\theta &= \frac{1}{2}\left(Pe_{f}S(z)\cos^{2}\theta\sin\phi\,\Psi-\frac{\partial \Psi}{\partial \theta} \right), \\
J_\phi&=\frac{1}{2}\left(Pe_{f}S(z)\cos\theta\cos\phi\,\Psi-\frac{1}{\sin\theta}\frac{\partial \Psi}{\partial \phi}\right). \label{eq:J3}
\end{align}
Denoting by $\hat{\v{n}}({z},\theta)$ the normal unit vector on one of the two hypersurfaces, the generalized no-flux condition is simply expressed as
\begin{equation}
\hat{\v{n}}(z,\theta)\bcdot\v{J}(z,\v{p},\Psi)=0,
\end{equation} 
which, upon calculation of the normal $\hat{\v{n}}$, leads to the two conditions:
\begin{align}
&J_{z}\mp L^{*}\textcolor{black}{\sin\theta} J_{\theta}=0\qquad \mathrm{at}\,\,\, z=1-L^{*}\,|\hspace{-0.05cm}\cos\theta|,\label{eq:HS1} \\
&J_{z}\pm L^{*}\textcolor{black}{\sin\theta} J_{\theta}=0\qquad \mathrm{at}\,\,\, z=-1+L^{*}\,|\hspace{-0.05cm}\cos\theta|.\label{eq:HS2}
\end{align}
In each case, the upper sign is used when $\theta\in[0,\upi/2]$ and the lower one when $\theta\in[\upi/2,\upi]$. Numerical solution of the conservation equation (\ref{eq:goveq0}) subject to the boundary conditions (\ref{eq:HS1})--(\ref{eq:HS2}) can be done using finite volumes as described in Appendix B. Typical results for the concentration profile $c(z)$ in the absence of flow are shown in figure~\ref{fig:exclusion} for different values of $L^{*}$ and compared to the solution obtained previously using the boundary condition (\ref{eq:BC00}), which corresponds to the limit of $L^{*}\rightarrow 0$. When steric exclusion is accounted for, a depletion layer is observed close to the walls whose thickness is of the order of $L^{*}$. \textcolor{black}{Steric exclusion leads to a decrease in concentration in the near wall region because it suppresses the orientations aligned towards the wall and hence the wall normal polarization.} Under stronger confinement (higher $L^{*}$), this leads to a concentration peak at the edge of the depletion layer due to wall accumulation, and this peak increases in magnitude and shifts closer to the wall as $L^{*}$ decreases. For very small values of $L^{*}$, the concentration profile approaches the profile obtained by neglecting steric effects, and steric exclusion can be safely neglected outside of the depletion layer itself whenever $L^{*}\lesssim 0.01$. This is indeed the appropriate regime in most microfluidic experiments with bacterial suspensions, which justifies the use of the simpler boundary condition (\ref{eq:BC00}) in the  work presented here.

\section{Finite-volume numerical algorithm}

In this Appendix, we describe the algorithm used for the numerical solution of equation (\ref{eq:goveq1}) for the distribution function. The method  is based on a finite-volume discretization of the Smoluchowski equation \citep{Ferziger02}, which has the advantage of satisfying conservation locally to machine precision while also allowing for an easy implementation of no-flux boundary conditions such as (\ref{eq:BC00}) or (\ref{eq:HS1})--(\ref{eq:HS2}). To avoid the cost of large matrix inversions, we solve the time-dependent Smoluchowski equation to steady state using an explicit scheme. In conservative form, the governing equation can be written as
\begin{equation}
\frac{\partial \Psi}{\partial t}+\bnabla_{J}\bcdot\v{J}=0, \label{eq:timefp}
\end{equation}
where $\v{J}$ is the generalized flux vector defined in equations (\ref{eq:J0})--(\ref{eq:J3}), and $\bnabla_{J}$ is the gradient operator in the three-dimensional $(z,\theta,\phi)$ space of particle configurations:
\begin{equation}
\bnabla_J\equiv\frac{\partial}{\partial z}\hat{\v{z}}+\frac{\partial}{\partial\theta}\hat{\boldsymbol{\theta}}+\frac{1}{\sin\theta}\frac{\partial}{\partial\phi}\hat{\boldsymbol{\phi}}.
\end{equation}
We note that $\Psi(z,\theta,\phi)$ is defined on a hypervolume obtained by extruding the unit sphere in the $z$ dimension. This computational domain is discretized into finite volumes using a uniform grid with respect to $(z,r,\phi)$, where $r=\cos\theta$. The nodal points $(z^{i},r^{j},\phi^{k})$  where $\Psi$ is evaluated are located at the centers of each volume and have coordinates
\begin{align}
&z^{i}=\frac{2i-1}{N_z}-1 \quad \,\,\,\,\,\,\,\,\,\mathrm{for}\quad i=1,...,N_{z}, \\
&r^{j}=\frac{2j-1}{N_r}-1 \quad \,\,\,\,\,\,\,\,\mathrm{for}\quad j=1,...,N_{r}, \\
&\phi^{k}=\frac{2\upi(k-1)}{N_\phi} \,\,\,\quad\quad \mathrm{for}\quad k=1,...,N_{\phi},
\end{align}
where $N_{z}$, $N_{r}$, and $N_{\phi}$ are the total numbers of points in each direction. We also define the grid spacing in each direction as
\begin{equation}
\Delta z = \frac{2}{N_{z}}, \quad \Delta r = \frac{2}{N_{r}}, \quad \Delta \phi =\frac{2\upi}{N_{\phi}}.
\end{equation}
The advantage of this discretization (compared to a uniform grid with respect to $\theta$) is that it divides the sphere of orientations into elements of equal area, which reduces restrictions on the time step arising from the rotational flux. 

\begin{figure}
\centering\vspace{0.0cm}
\includegraphics[width=11.5cm]{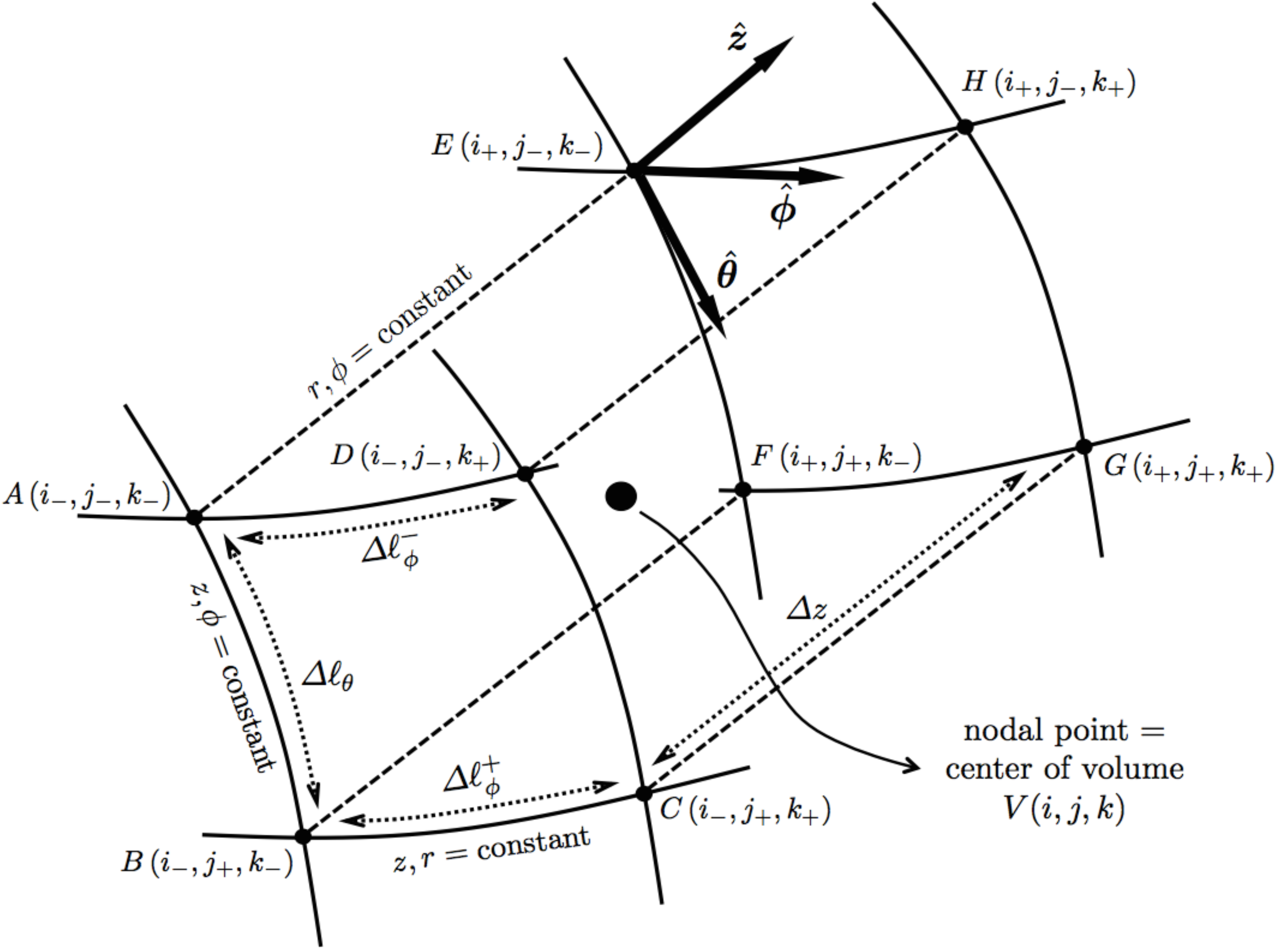}
\caption{Typical finite volume in three-dimensional $(z,\theta,\phi)$ space, centered around an arbitrary nodal point with indices $(i,j,k)$. The uniform discretization with respect to $(z,r,\phi)$ ensures that all such computational cells have equal volume $\Delta V=\Delta z\Delta r\Delta \phi$.  } \label{fig:finitevolume}
\end{figure}

A typical finite volume centered around node $(i,j,k)$ is illustrated in figure~\ref{fig:finitevolume}. It is delimited by eight grid points denoted $A$ through $H$, with indices $(i_{\pm},j_{\pm},k_{\pm})$ where we have  introduced the notations $i_{\pm}=i\pm0.5$, $j_{\pm}=j\pm0.5$ and $k_{\pm}=k\pm0.5$. The cell edges have lengths \vspace{-0.05cm}
\begin{align}
&AB=DC=EF=HG=\Delta\ell_{\theta}\equiv\cos^{-1}(r^{j_{-}})-\cos^{-1}(r^{j_{+}}), \\
&AD=EH=\Delta\ell_{\phi}^{-}\equiv \frac{2\upi \sin \theta^{j_{-}}}{N_{\phi}}, \quad     BC=FG=\Delta\ell_{\phi}^{+}\equiv\frac{2\upi \sin \theta^{j_{+}}}{N_{\phi}}, \\
&AE=BF=DH=CG=\Delta z.\vspace{-0.05cm}
\end{align}
In figure~\ref{fig:finitevolume}, faces $ABCD$ and $EFGH$ have unit normal $\hat{\v{z}}$ and surface area $\Delta r\Delta \phi$. Similarly, faces $ADHE$ and $BCGF$ have unit normal $\hat{\boldsymbol{\theta}}$ and areas $\Delta z\Delta \ell_{\phi}^{-}$ and $\Delta z\Delta \ell_{\phi}^{+}$, respectively, whereas faces $ABFE$ and $DCGH$ have unit normal $\hat{\boldsymbol{\phi}}$ and area $\Delta z\Delta\ell_{\theta}$. The volume of the computational cell is  $\Delta V=\Delta z\Delta r\Delta \phi$. 

In order to satisfy conservation of the distribution function exactly in each finite volume, we first integrate equation (\ref{eq:timefp}) over  computational cell $V(i,j,k)$:
\begin{equation}
\iiint\limits_{V(i,j,k)}\left(\frac{\partial \Psi}{\partial t}+\bnabla_{J}\bcdot\v{J}\right)\,\mathrm{d}z\,\mathrm{d}r\,\mathrm{d}\phi=0. 
\end{equation}
After applying the divergence theorem to the second term, this can be recast as
\begin{align}
\begin{split}
0=\frac{\partial}{\partial t}\iiint\limits_{V(i,j,k)}\Psi\,\mathrm{d}z\,\mathrm{d}r\,\mathrm{d}\phi&+\iint\limits_{ABCD}J_{z}\,\mathrm{d}r\,\mathrm{d}\phi-\iint\limits_{EFGH}J_{z}\,\mathrm{d}r\,\mathrm{d}\phi \\
&+\iint\limits_{ADHE}J_{\theta}\,\mathrm{d}z\,\mathrm{d}\phi-\iint\limits_{BCGF}J_{\theta}\,\mathrm{d}z\,\mathrm{d}\phi \\
&+\iint\limits_{ABFE}J_{\phi}\,\mathrm{d}z\,\mathrm{d}r-\iint\limits_{DCGH}J_{\phi}\,\mathrm{d}z\,\mathrm{d}r.
\end{split} \label{eq:fv1}
\end{align}
Volume and surface integrals in equation (\ref{eq:fv1}) are approximated to second-order using a midpoint rule. After division by $\Delta V$, this leads to the discretized equation:
\begin{align}
\begin{split}
0=\frac{\partial \Psi^{i,j,k}}{\partial t}&+\frac{1}{\Delta z}\left[J_{z}(i_{+},j,k)-J_{z}(i_{-},j,k)\right] \\
&+\frac{1}{\Delta r}\left[J_{\theta}(i,j_{+},k)\sin\theta^{j_{+}}-J_{\theta}(i,j_{-},k)\sin\theta^{j_{-}}\right] \\
&+\frac{\Delta\ell_{\theta}}{\Delta r\Delta \phi}\left[J_{\phi}(i,j,k_{+})-J_{\phi}(i,j,k_{-})\right]. \label{eq:discrete}
\end{split}
\end{align}
In order to integrate this equation, we must first obtain approximate expressions for the fluxes at the centers of the six volume faces. This is done using linear interpolation for terms involving $\Psi$, and centered finite differences for terms involving derivatives of $\Psi$. In the $z$ and $\phi$ directions, this gives
\begin{align}
&J_{z}(i_{+},j,k)\approx Pe_{s}\cos\theta^{j} \left(\frac{\Psi^{i+1,j,k}+\Psi^{i,j,k}}{2}\right)-\textcolor{black}{2\Lambda}Pe^{2}_{s} \left(\frac{\Psi^{i+1,j,k}-\Psi^{i,j,k}}{\Delta z}\right), \\
\begin{split}
&J_{\phi}(i,j,k_{+})\approx \frac{1}{2}\left[Pe_{f}S(z^{i})\cos\theta^{j}\cos\phi^{k_{+}}\left(\frac{\Psi^{i,j,k+1}+\Psi^{i,j,k}}{2}\right)\right. \\
& \qquad \qquad \qquad  \qquad \qquad \left.-\frac{1}{\sin\theta^{j}}\left(\frac{\Psi^{i,j,k+1}-\Psi^{i,j,k}}{\Delta \phi}\right) \right],
\end{split}
\end{align}
with similar expressions for $J_{z}(i_{-},j,k)$ and $J_{\phi}(i,j,k_{-})$. 
The approximation of  $J_{\theta}$ is slightly more involved due to the non-uniformity of the mesh with respect to $\theta$. Derivatives with respect to $\theta$ are calculated using symmetric central finite differences in terms of $r$ after application of the chain rule, and linear interpolation is used with respect to the $\theta$ variable, leading to the approximation
\begin{align}
\begin{split}
&J_{\theta}(i,j_{+},k)\approx \frac{1}{2}\left\{Pe_{f}S(z^{i})\cos\theta^{j_{+}}\cos\phi^{k}\left[\lambda^{j_{+}}\Psi^{i,j+1,k}+(1-\lambda^{j_{+}})\Psi^{i,j,k}\right]\right. \\
&\qquad\qquad\qquad\qquad\qquad\qquad+\left.\sin\theta^{j_{+}}\left(\frac{\Psi^{i,j+1,k}-\Psi^{i,j-1,k}}{\Delta r}\right)\right\},
\end{split}
\end{align}
with a similar expression for $J_{\theta}(i,j_{-},k)$. The interpolation weight $\lambda^{j_{+}}$ is given by
\begin{equation}
\lambda^{j_{+}}=\frac{\cos^{-1}(r^{j}+\frac{\Delta r}{2})-\cos^{-1}(r^{j})}{\cos^{-1}(r_{j}+\Delta r)-\cos^{-1}(r^{j})}.
\end{equation}
When integrating equation (\ref{eq:discrete}) in time, care must be taken when dealing with cells adjacent to the poles of the unit sphere ($j=1$ and $N_{r}$), as these cells are missing one face. For instance, cells with $j=1$ are such that $A=D$ and $E=H$ in the diagram of figure~\ref{fig:finitevolume}, so that face $ADHE$ is missing and the corresponding flux should not be included in the discretized equation.  

Boundary conditions also need to be specified to proceed with the time integration. Periodic boundary conditions are used in the $\phi$ direction, yielding: 
\begin{equation}
J_{\phi}(1/2,j,k)=J_{\phi}(N_{\phi}-1/2,j,k) \quad \mathrm{and} \quad J_{\phi}(N_{\phi}+1/2,j,k)=J_{\phi}(3/2,j,k).
\end{equation}
Treatment of the boundaries in the $\theta$ and $z$ directions differs depending on whether steric exclusion with the walls is included or not.

\subsection{Without steric exclusion}

When steric exclusion is not included and the simple boundary condition of equation (\ref{eq:BC00}) is used, $\theta$ varies over its full range $[0,\upi]$. However, no boundary condition is needed along $\theta$ as the boundary cells with $j=1$ and $N_{r}$ are missing one face as explained above, which eliminates the need to specify $J_{\theta}(i,1/2,k)$ and $J_{\phi}(i,N_{r}+1/2,k)$. Along the $z$ direction, the boundary condition is simply the no-flux condition (\ref{eq:BC00}), which translates into
\begin{equation}
J_{z}(i,j,1/2)=J_z(i,j,N_{z}+1/2)=0.
\end{equation}

\subsection{With steric exclusion}

The situation is more complex when steric exclusion is accounted for, as the boundary conditions needs to be enforced on the hypersurfaces defined in equations (\ref{eq:hyper1})--(\ref{eq:hyper2}). It is convenient in this case to choose $N_{z}$ and $N_{r}$ such that
\begin{equation}
\Delta z = L^{*} \Delta r \quad \mathrm{or} \quad N_{z}=\frac{N_{r}}{L^{*}}. \label{eq:NzNr}
\end{equation}
Indeed this ensures that the hypersurfaces fall onto grid points and eliminates the need for further interpolation. However, if $L^{*}$ is small, this implies that a significantly finer resolution is needed along $z$ than along $\theta$. As we discussed in Appendix A, the hypersurfaces limit the range of allowable values of $\theta$ to an interval of the form $[\theta_{1}(z),\theta_{2}(z)]\subset[0,\upi]$ for particles located near the walls. After discretization of the domain and choosing $N_{z}$ and $N_{r}$ to satisfy condition (\ref{eq:NzNr}), we find that for  any nodal point with coordinate $z^{i}$, there is a finite range $[\theta^{j_{1}(i)},\theta^{j_{2}(i)}]$ of allowable values of $\theta^{j}$, with
\begin{eqnarray}
&&j_{1}(i) = \begin{cases} \displaystyle \frac{N_{r}}{2}+1-i & \text{if $z\leq -1+L^{*}$,}\vspace{0.1cm}  \\ \displaystyle\frac{N_{r}}{2}-N_{z}+i & \text{if $z\geq 1-L^{*}$,} \vspace{0.1cm} \\ \,\,1 & \text{otherwise,} \end{cases}\\
&&j_{2}(i) = \begin{cases} \displaystyle\frac{N_{r}}{2}+i & \text{if $z\leq -1+L^{*}$,} \vspace{0.1cm} \\ \displaystyle\frac{N_{r}}{2}+N_{z}+1-i & \text{if $z\geq 1-L^{*}$,}  \vspace{0.1cm}\\ \,\,N_{r} & \text{otherwise.} \end{cases}
\end{eqnarray}
Interior nodal points such that $j\in[j_{1}(i)+1,j_{2}(i)-1]$ are such that full cuboidal finite volumes in $(z,r,\phi)$ can be constructed around them, and therefore do not require any special boundary treatment. Boundary nodal points such that $j=j_{1}(i)$ or $j_{2}(i)$, however, are contained inside prisms whose hypotenuses  coincide with the hypersurfaces. These finite volumes can be treated in the same way as interior control volumes by prescribing zero-flux contributions from surfaces lying outside of the domain, by multiplying the volume $\Delta V$ by 0.5, and by adjusting the surface area of faces $ABFE$ and $DCGH$ to a reduced triangular area given by
\begin{eqnarray}
\Delta A  = \frac{\Delta z \Delta l_{\theta}}{2} + \left[r^{j}\Delta l_{\theta} - 2 \sin\left(\frac{\Delta l_{\theta}}{2}\right)\cos\left(\frac{\theta^{j_{+}}+\theta^{j_{-}}}{2}\right)\right].
\end{eqnarray}

\section{Active particle trajectories and shear trapping}

In this Appendix, we rationalize the linear dependence of the depletion layer thickness $\delta_{D}$ upon $Pe_{s}/Pe_{f}$ by deriving the trajectory of a deterministic swimmer whose dynamics result from self-propulsion and shear rotation via Jeffery's equation. A similar derivation was previously presented by \cite{Zottl12,Zottl13}. In dimensional variables, the equations of motion of the swimmer are written
\begin{align}
\dot{z}(t)&=V_{s}\cos\theta(t), \label{eq:zdot} \\
\dot{\v{p}}(t)&=(\t{I}-\v{pp})\bcdot(\zeta \t{E}+\t{W})\bcdot\v{p}. \label{eq:jeffery}
\end{align}
Here, $\zeta$ is a shape parameter, with $\zeta\approx 1$ for a slender particle as we have assumed in the rest of the paper. The two second-order tensors $\t{E}$ and $\t{W}$ are the rate-of-strain and vorticity tensors of the imposed flow, respectively:
\begin{equation}
\t{E}=\frac{\dot{\gamma}_{w}}{2}z(t)\left(\hat{\v{y}}\hat{\v{z}} + \hat{\v{z}}\hat{\v{y}}\right), \qquad \t{W}=\frac{\dot{\gamma}_{w}}{2}z(t)\left(\hat{\v{y}}\hat{\v{z}} - \hat{\v{z}}\hat{\v{y}}\right).
\end{equation}
Parameterizing the orientation vector as $\v{p}=(\sin\theta\cos\phi,\sin\theta\sin\phi, \cos\theta)$, we can use equation (\ref{eq:jeffery}) to obtain expressions for the time rate of change of the polar and azimuthal angles of the swimmer as
\begin{align}
\dot{\theta}(t) &= \frac{\dot{\gamma}_{w}}{2}z(t)\textcolor{black}{\sin\phi(t)}\left[\left(\zeta+1\right)\cos^2\theta(t) - \left(\zeta -1\right) \sin^2\theta(t)\right], \label{eq:thetadot}\\ 
\dot{\phi}(t) &= \frac{\dot{\gamma}_{w}}{2}z(t)\left(\zeta+1\right) \frac{\cos\theta(t)\textcolor{black}{\cos\phi(t)}}{\sin\theta(t)}. \label{eq:phidot}
\end{align}
Equations (\ref{eq:zdot}), (\ref{eq:thetadot}) and (\ref{eq:phidot}) form a closed system of coupled ordinary differential equations that can be solved for the swimmer dynamics. 

Any swimmer that is not perfectly aligned with the walls ($\cos\theta \ne 0$) will tend to migrate towards one of the boundaries due to self-propulsion, while shear rotation tends to align it along the flow direction causing it to get trapped. Recalling the definition of  $\chi$ as the ratio of the time scale for shear rotation to the time it takes for a swimmer to cross the channel,
\begin{equation}
\chi = \frac{V_{s}}{2\dot{\gamma}_{w}H}=\frac{Pe_{s}}{Pe_{f}}, 
\end{equation}
we expect two different regimes.
When $\chi \gg 1$, any swimmer released from the centerline with initial orientation $(\theta_0,\phi_0)$ will reach one of the walls before becoming trapped. On the other hand, when $\chi \ll 1$, we expect there to exist a position $z_{trap}(\theta_{0},\phi_{0})$ inside the channel where the swimmer gets trapped due to shear alignment. This indeed corresponds to the regime discussed in \S \ref{sec:strongflow}, where depletion from the centerline and shear-trapping were predicted to occur for $Pe_{s}\ll 1$ and $Pe_f\gg 1$. 

To derive a quantitative estimate for $z_{trap}$, we calculate the value of $z$ at which $\theta$ first reaches $\pm \upi/2$. We first consider the case of a particle with initial position $z_{0}=0$ and orientation defined by $\theta_{0}\in[0,\upi/2)$, $\phi_0=\textcolor{black}{3\upi/2}$. For this specific initial configuration, $\dot{\phi}(0) = 0$ which implies $\phi(t) = \textcolor{black}{3\upi/2}$ for all times. The motion is two-dimensional in this case, and the dynamics is governed by the two coupled ordinary differential equations
\begin{align}
\dot{z}(t) &= V_{s} \cos\theta(t), \label{eq:zdot2}\\
\dot{\theta}(t) &= -\frac{\dot{\gamma_{w}}}{2}z(t)\left[\left(\zeta+1\right)\cos^2\theta(t) - \left(\zeta -1\right) \sin^2\theta(t)\right]. \label{eq:thetadot2}
\end{align}
An equation for the swimmer trajectory can then be obtained by taking the ratio of (\ref{eq:zdot2}) and (\ref{eq:thetadot2}):
\begin{equation}
\frac{\mathrm{d}\theta}{\mathrm{d}z} = \frac{z}{H}\left[\frac{\left(\zeta+1\right) - 2\left(\zeta -1\right) \sin^2\theta(t)}{2\chi\cos\theta}\right].
\end{equation}
This can be integrated from $(z,\theta)=(0,\theta_{0})$ to $(z_{trap},\upi/2)$, yielding 
\begin{equation}
\left(\frac{z_{trap}(\theta_0)}{H}\right)^2 = \chi\sqrt{\frac{1}{2\zeta(\zeta + 1)}} \left(\tanh^{-1}\sqrt{\frac{2\zeta}{\zeta + 1}}- \tanh^{-1}\sqrt{\frac{2\zeta}{\zeta + 1}\sin\theta_{0}}\right).\label{eq:ztrap}
\end{equation}
For a typical swimmer of aspect ratio $10$, we estimate $\zeta\approx 0.98$. Taking the initial configuration to be $\theta_0=0$, equation (\ref{eq:ztrap}) simplifies to $z_{trap}/H\approx \sqrt{3\chi}\approx 1.73 \sqrt{Pe_s/Pe_f}$. This estimate is consistent with the high-$Pe_f$ scaling analysis of \S \ref{sec:strongflow}, as well as with the numerical results of \S \ref{sec:trapping} where we found $\delta_{D}\approx 2.404 \sqrt{Pe_s/Pe_f}$.

The more general case of an arbitrary initial orientation $(\theta_0,\phi_0)$ can also be solved analytically. Combining equations (\ref{eq:thetadot}) and (\ref{eq:phidot}) to eliminate $z(t)$, we find after integration:\textcolor{black}{\begin{equation}
\cos\phi =\cos\phi_{0} \bigg| \frac{\left(\zeta+1 \right)\mathrm{cosec}^2\theta - 2\zeta}{\left(\zeta+1\right)\mathrm{cosec}^2\theta_{0}- 2\zeta}\bigg|^{\frac{1}{2}}. \label{eq:sinphi}
\end{equation}}
Now, using equations (\ref{eq:zdot}) and (\ref{eq:thetadot}), we get 
\textcolor{black}{
\begin{equation}
\left(\frac{z_{trap}(\phi_{0},\theta_{0})}{H}\right)^{2} = 2\chi\int_{\theta_{0}}^{\pi/2} \frac{\cos\theta}{\left(\zeta + 1 - 2\,\zeta\sin^2\theta\right) \sqrt{1-\cos^{2}\phi}} \,\mathrm{d}\theta, \label{eq:ztrap2}
\end{equation}}where $\sin\phi$ is known in terms of $\theta$ using (\ref{eq:sinphi}). This expression confirms the scaling of $z_{trap}$ with $\sqrt{\chi}$, and it can in fact be shown that $z_{trap}$ in equation (\ref{eq:ztrap2}) has an upper bound given by the previous estimate  (\ref{eq:ztrap}).

\bibliographystyle{jfm}
\bibliography{BigBib}

\begin{thebibliography}{73}
\expandafter\ifx\csname natexlab\endcsname\relax\def\natexlab#1{#1}\fi

\bibitem[Altshuler {\em et~al.\/}(2013)Altshuler, {Mi\~no}, {P\'erez-Penichet},
  {del R\'io}, Lindner, Rousselet \& {Cl\'ement}]{Altshuler13}
{\sc Altshuler, E., {Mi\~no}, G., {P\'erez-Penichet}, C., {del R\'io}, L.,
  Lindner, A., Rousselet, A. \& {Cl\'ement}, E.} 2013 Flow-controlled
  densification and anomalous dispersion of \textit{E. coli} through a
  constriction. {\em Soft Matter\/} {\bf 9}, 1864--1870.

\bibitem[Baskaran \& Marchetti(2009)]{Baskaran09}
{\sc Baskaran, A. \& Marchetti, M.~C.} 2009 Statistical mechanics and
  hydrodynamics of bacterial suspensions. {\em Proc. Natl. Acad. Sci. USA\/}
  {\bf 106}, 15567--15572.

\bibitem[Bearon {\em et~al.\/}(2011)Bearon, Hazel \& Thorn]{Bearon11}
{\sc Bearon, R.~N., Hazel, A.~L. \& Thorn, G.~J.} 2011 The spatial distribution
  of gyrotactic swimming micro-organisms in laminar flow fields. {\em J. Fluid
  Mech.\/} {\bf 680}, 602--635.

\bibitem[Berke {\em et~al.\/}(2008)Berke, Turner, Berg \& Lauga]{Berke08}
{\sc Berke, A.~P., Turner, L., Berg, H.~C. \& Lauga, E.} 2008 Hydrodynamic
  attraction of swimming microorganisms by surfaces. {\em Phys. Rev. Lett.\/}
  {\bf 101}, 038102.

\bibitem[Bretherton(1962)]{Bretherton62}
{\sc Bretherton, F.~P.} 1962 The motion of rigid particles in a shear flow at
  low {Reynolds} number. {\em J. Fluid Mech.\/} {\bf 14}, 284--304.

\bibitem[Cellia {\em et~al.\/}(2009)Cellia, Turner, Afdhal, Keates, Ghiran,
  Kelly, Ewoldt, {McKinley}, So, Erramilli \& Bansil]{Cellia09}
{\sc Cellia, J.~P., Turner, B.~S., Afdhal, N.~H., Keates, S., Ghiran, I.,
  Kelly, C.~P., Ewoldt, R.~H., {McKinley}, G.~H., So, P., Erramilli, S. \&
  Bansil, R.} 2009 \textit{Helicobacter pylori} moves through mucus by reducing
  mucin viscoelasticity. {\em Proc. Natl. Acad. Sci. USA\/} {\bf 106},
  14321--14326.

\bibitem[Chilukuri {\em et~al.\/}(2014)Chilukuri, Collins \&
  Underhill]{Chilukuri14}
{\sc Chilukuri, S., Collins, C.~H. \& Underhill, P.~T.} 2014 Impact of external
  flow on the dynamics of swimming microorganisms near surfaces. {\em J. Phys.:
  Condens. Matter\/} {\bf 26}, 115101.

\bibitem[Costanzo {\em et~al.\/}(2012)Costanzo, {Di Leonardo}, Ruocco \&
  Angelani]{Costanzo12}
{\sc Costanzo, A., {Di Leonardo}, R., Ruocco, G. \& Angelani, L.} 2012
  Transport of self-propelling bacteria in micro-channel flow. {\em J. Phys.:
  Condens. Matter\/} {\bf 24}, 065101.

\bibitem[Denissenko {\em et~al.\/}(2012)Denissenko, Kanstler, Smith \&
  {Kirkman-Brown}]{Denissenko12}
{\sc Denissenko, P., Kanstler, V., Smith, D.~J. \& {Kirkman-Brown}, J.} 2012
  Human spermatozoa migration in micro channels reveals boundary-following
  navigation. {\em Proc. Natl. Acad. Sci. USA\/} {\bf 109}, 8007--8010.

\bibitem[{Di Leonardo} {\em et~al.\/}(2010){Di Leonardo}, Angelani,
  {Dell'Arciprete}, Ruocco, Iebba, Schippa, Conte, Mecarini, {De Angelis} \&
  {Di Fabrizio}]{Dileonardo10}
{\sc {Di Leonardo}, R., Angelani, L., {Dell'Arciprete}, D., Ruocco, G., Iebba,
  V., Schippa, S., Conte, M.~P., Mecarini, F., {De Angelis}, F. \& {Di
  Fabrizio}, E.} 2010 Bacterial ratchet motors. {\em Proc. Natl. Acad. Sci.
  USA\/} {\bf 107}, 9541--9545.

\bibitem[Doi \& Edwards(1986)]{Doi86}
{\sc Doi, M. \& Edwards, S.~F.} 1986 {\em The Theory of Polymer Dynamics\/}.
  Oxford University Press.

\bibitem[Drescher {\em et~al.\/}(2011)Drescher, Dunkel, Cisneros, Ganguly \&
  Goldstein]{Drescher11}
{\sc Drescher, K., Dunkel, J., Cisneros, L.~H., Ganguly, S. \& Goldstein,
  R.~E.} 2011 Fluid dynamics and noise in bacterial cell-cell and cell-surface
  scattering. {\em Proc. Natl. Acad. Sci. USA\/} {\bf 108}, 10940--10945.

\bibitem[Edwards \& Yeomans(2009)]{Edwards09}
{\sc Edwards, S.~A. \& Yeomans, J.~M.} 2009 Spontaneous flow states in active
  nematics: a unified picture. {\em Europhys. Lett.\/} {\bf 85}, 18008.

\bibitem[Elgeti \& Gompper(2013)]{Elgeti13}
{\sc Elgeti, J. \& Gompper, G.} 2013 Wall accumulation of self-propelled
  spheres. {\em Europhys. Lett.\/} {\bf 101}, 48003.

\bibitem[Elgeti \& Gompper(2015)]{Elgeti15}
{\sc Elgeti, J. \& Gompper, G.} 2015 Run-and-tumble dynamics of self-propelled
  particles in confinement. {\em Europhys. Lett.\/} {\bf 109}, 58003.

\bibitem[Ezhilan {\em et~al.\/}(2012)Ezhilan, Pahlavan \&
  Saintillan]{Ezhilan12}
{\sc Ezhilan, B., Pahlavan, A.~A. \& Saintillan, D.} 2012 Chaotic dynamics and
  oxygen transport in thin films of aerotactic bacteria. {\em Phys. Fluids\/}
  {\bf 24}, 091701.

\bibitem[Fauci \& {McDonald}(1995)]{Fauci95}
{\sc Fauci, L.~J. \& {McDonald}, A.} 1995 Sperm motility in the presence of
  boundaries. {\em Bull. Math. Biol.\/} {\bf 57}, 679--699.

\bibitem[Ferziger \& {Peri\'c}(2002)]{Ferziger02}
{\sc Ferziger, J.~H. \& {Peri\'c}, M.} 2002 {\em Computational Methods for
  Fluid Dynamics\/}. Springer.

\bibitem[Forest {\em et~al.\/}(2013)Forest, Wang \& Zhou]{Forest13}
{\sc Forest, M.~G., Wang, Q. \& Zhou, R.} 2013 Kinetic theory and simulations
  of active polar liquid crystalline polymers. {\em Soft Matter\/} {\bf 9},
  5207--5222.

\bibitem[F\"urthauer {\em et~al.\/}(2012)F\"urthauer, Neef, Grill, Kruse \&
  J\"ulicher]{Furthauer12}
{\sc F\"urthauer, S., Neef, M., Grill, S.~W., Kruse, K. \& J\"ulicher, F.} 2012
  The {Taylor-Couette} motor: spontaneous flows of active polar fluids between
  two coaxial cylinders. {\em New J. Phys.\/} {\bf 14}, 023001.

\bibitem[Gachelin {\em et~al.\/}(2014)Gachelin, Rousselet, Lindner \&
  Clement]{Gachelin14}
{\sc Gachelin, J., Rousselet, A., Lindner, A. \& Clement, E.} 2014 Collective
  motion in an active suspension of \textit{Escherichia coli} bacteria. {\em
  New J. Phys.\/} {\bf 16}, 025003.

\bibitem[Galajda {\em et~al.\/}(2007)Galajda, Keymer, Chaikin \&
  Austin]{Galajda07}
{\sc Galajda, P., Keymer, J., Chaikin, P. \& Austin, R.} 2007 A wall of funnel
  concentrates swimming bacteria. {\em J. Bacteriol.\/} {\bf 189}, 8704--8707.

\bibitem[Garcia {\em et~al.\/}(2011)Garcia, Berti, Peyla \&
  {Rafa\"i}]{Garcia11}
{\sc Garcia, M., Berti, S., Peyla, P. \& {Rafa\"i}, S.} 2011 Random walk of a
  swimmer in a low-{Reynolds}-number medium. {\em Phys. Rev. E\/} {\bf 83},
  035301.

\bibitem[Gibbs {\em et~al.\/}(2011)Gibbs, Kothari, Saintillan \& Zhao]{Gibbs11}
{\sc Gibbs, J.~G., Kothari, S., Saintillan, D. \& Zhao, Y.-P.} 2011
  Geometrically designing the kinematic behavior of catalytic nanomotors. {\em
  Nano Lett.\/} {\bf 11}, 2543--2550.

\bibitem[{Hern\'andez-Ortiz} {\em et~al.\/}(2005){Hern\'andez-Ortiz}, Stoltz \&
  Graham]{Hernandez05}
{\sc {Hern\'andez-Ortiz}, J.~P., Stoltz, C.~G. \& Graham, M.~D.} 2005 Transport
  and collective dynamics in suspensions of confined swimming particles. {\em
  Phys. Rev. Lett.\/} {\bf 95}, 204501.

\bibitem[{Hern\'andez-Ortiz} {\em et~al.\/}(2009){Hern\'andez-Ortiz}, Underhill
  \& Graham]{Hernandez09}
{\sc {Hern\'andez-Ortiz}, J.~P., Underhill, P.~T. \& Graham, M.~D.} 2009
  Dynamics of confined suspensions of swimming particles. {\em J. Phys.:
  Condens. Matter\/} {\bf 21}, 204107.

\bibitem[Hill {\em et~al.\/}(2007)Hill, Kalkanci, {McMurry} \& Koser]{Hill07}
{\sc Hill, J., Kalkanci, O., {McMurry}, J.~L. \& Koser, H.} 2007 Hydrodynamic
  surface interactions enable \textit{Escherichia coli} to seek efficient
  routes to swim upstream. {\em Phys. Rev. Lett.\/} {\bf 98}, 068101.

\bibitem[Hulme {\em et~al.\/}(2008)Hulme, {DiLuzio}, Shevkoplyas, Turner,
  Mayer, Berg \& Whitesides]{Hulme08}
{\sc Hulme, S.~E., {DiLuzio}, W.~R., Shevkoplyas, S.~S., Turner, L., Mayer, M.,
  Berg, H.~C. \& Whitesides, G.~M.} 2008 Using ratchets and sorters to
  fractionate motile cells of \textit{Escherichia coli} by length. {\em Lab on
  a Chip\/} {\bf 8}, 1888--1895.

\bibitem[Jeffery(1922)]{Jeffery22}
{\sc Jeffery, G.~B.} 1922 The motion of ellipsoidal particles immersed in a
  viscous fluid. {\em Proc. R. Soc. Lond. A\/} {\bf 102}, 161--179.

\bibitem[Kaiser {\em et~al.\/}(2014)Kaiser, Peshkov, Sokolov, {ten Hagen},
  L\"owen \& Aranson]{Kaiser14}
{\sc Kaiser, A., Peshkov, A., Sokolov, A., {ten Hagen}, B., L\"owen, H. \&
  Aranson, I.~S.} 2014 Transport powered by bacterial turbulence. {\em Phys.
  Rev. Lett.\/} {\bf 112}, 158101.

\bibitem[Kaiser {\em et~al.\/}(2012)Kaiser, Wensink \& {L\"owen}]{Kaiser12}
{\sc Kaiser, A., Wensink, H.~H. \& {L\"owen}, H.} 2012 How to capture active
  particles. {\em Phys. Rev. Lett.\/} {\bf 108}, 268307.

\bibitem[Kantsler {\em et~al.\/}(2014)Kantsler, Dunkel, Blayney \&
  Goldstein]{Kantsler14}
{\sc Kantsler, V., Dunkel, J., Blayney, M. \& Goldstein, R.~E.} 2014 Rheotaxis
  facilitates upstream navigation of mammalian sperm cells. {\em eLife\/} {\bf
  3}, 02403.

\bibitem[Kantsler {\em et~al.\/}(2013)Kantsler, Dunkel, Polin \&
  Goldstein]{Kantsler13}
{\sc Kantsler, V., Dunkel, J., Polin, M. \& Goldstein, R.~E.} 2013 Ciliary
  contact interactions dominate surface scattering of swimming eukaryotes. {\em
  Proc. Natl. Acad. Sci. USA\/} {\bf 110}, 1187--1192.

\bibitem[Kasyap \& Koch(2014)]{Kasyap14a}
{\sc Kasyap, T.~V. \& Koch, D.} 2014 Instability of an inhomogeneous bacterial
  suspension subjected to a chemo-attractant gradient. {\em J. Fluid Mech.\/}
  {\bf 741}, 619--657.

\bibitem[Kaya \& Koser(2009)]{Kaya09}
{\sc Kaya, T. \& Koser, H.} 2009 Characterization of hydrodynamic surface
  interactions of \textit{Escherichia coli} cell bodies in shear flow. {\em
  Phys. Rev. Lett.\/} {\bf 103}, 138103.

\bibitem[Kaya \& Koser(2012)]{Kaya12}
{\sc Kaya, T. \& Koser, H.} 2012 Direct upstream motility in
  \textit{Escherichia coli}. {\em Biophys. J.\/} {\bf 102}, 1514--1523.

\bibitem[Kim {\em et~al.\/}(2014)Kim, Drescher, Park, Bassler \& Stone]{Kim14}
{\sc Kim, M.~Y., Drescher, K., Park, O.~S., Bassler, B. \& Stone, H.~A.} 2014
  Filaments in curved streamlines: rapid formation of \textit{Staphylococcus
  aureus} biofilm streamers. {\em N. J. Phys.\/} {\bf 16}, 065024.

\bibitem[Koumakis {\em et~al.\/}(2013)Koumakis, Lepore, Maggi \& {Di
  Leonardo}]{Koumakis13}
{\sc Koumakis, N., Lepore, A., Maggi, C. \& {Di Leonardo}, R.} 2013 Targeted
  delivery of colloids by swimming bacteria. {\em Nature Comm.\/} {\bf 4},
  2588.

\bibitem[Krochak {\em et~al.\/}(2010)Krochak, Olson \& Martinez]{Krochak10}
{\sc Krochak, P.~J., Olson, J.~A. \& Martinez, D.~M.} 2010 Near-wall estimates
  of the concentration and orientation distribution of a semi-dilute rigid
  fibre suspension in {Poiseuille} flow. {\em J. Fluid Mech.\/} {\bf 653},
  431--462.

\bibitem[Lambert {\em et~al.\/}(2010)Lambert, Liao \& Austin]{Lambert10}
{\sc Lambert, G., Liao, D. \& Austin, R.~H.} 2010 Collective escape of
  chemotactic swimmers through microscopic ratchets. {\em Phys. Rev. Lett.\/}
  {\bf 104}, 168102.

\bibitem[Lauga {\em et~al.\/}(2006)Lauga, {DiLuzio}, Whitesides \&
  Stone]{Lauga06}
{\sc Lauga, E., {DiLuzio}, W.~R., Whitesides, G.~M. \& Stone, H.~A.} 2006
  Swimming in circles: Motion of bacteria near solid boundaries. {\em Biophys.
  J.\/} {\bf 90}, 400--412.

\bibitem[Lecuyer {\em et~al.\/}(2011)Lecuyer, Rusconi, Chen, Forsyth, Vlamakis,
  Kolter \& Stone]{Lecuyer11}
{\sc Lecuyer, S., Rusconi, R., Chen, Y., Forsyth, A., Vlamakis, H., Kolter, R.
  \& Stone, H.~A.} 2011 Shear stress increases the residence time of adhesion
  of \textit{Pseudomonas aeruginosa}. {\em Biophys. J.\/} {\bf 100}, 341--350.

\bibitem[Lee(2013)]{Lee2013}
{\sc Lee, C.~F.} 2013 Active particles under confinement: aggregation at the
  wall and gradient formation inside a channel. {\em New J. Phys.\/} {\bf 15},
  055007.

\bibitem[Li \& Ardekani(2014)]{Li14}
{\sc Li, G. \& Ardekani, A.~M.} 2014 Hydrodynamic interaction of microswimmers
  near a wall. {\em Phys. Rev. E\/} {\bf 90}, 013010.

\bibitem[Li {\em et~al.\/}(2011)Li, Bensson, Nisimova, Munger, Mahautmr, Tang,
  Maxey \& Brun]{Li11}
{\sc Li, G., Bensson, J., Nisimova, L., Munger, D., Mahautmr, P., Tang, J.~X.,
  Maxey, M.~R. \& Brun, Y.~V.} 2011 Accumulation of swimming bacteria near a
  solid surface. {\em Phys. Rev. E\/} {\bf 84}, 041932.

\bibitem[Li \& Tang(2009)]{Li09}
{\sc Li, G. \& Tang, J.~X.} 2009 Accumulation of microswimmers near a surface
  mediated by collision and rotational {Brownian} motion. {\em Phys. Rev.
  Lett.\/} {\bf 103}, 078101.

\bibitem[Lu \& Walker(2001)]{Lu01}
{\sc Lu, L. \& Walker, W.~A.} 2001 Pathologic and physiologic interactions of
  bacteria with the gastrointestinal epithelium. {\em Am. J. Clin. Nutr.\/}
  {\bf 73}, 1124--1130.

\bibitem[Lushi {\em et~al.\/}(2014)Lushi, Wioland \& Goldstein]{Lushi14}
{\sc Lushi, E., Wioland, H. \& Goldstein, R.~E.} 2014 Fluid flows created by
  swimming bacteria drive self-organization in confined suspensions. {\em Proc.
  Natl. Acad. Sci. USA\/} {\bf 111}, 9733--9738.

\bibitem[Marchetti {\em et~al.\/}(2013)Marchetti, Joanny, Ramaswamy, Liverpool,
  Prost, Rao \& {Aditi Simha}]{Marchetti13}
{\sc Marchetti, M.~C., Joanny, J.~F., Ramaswamy, S., Liverpool, T.~B., Prost,
  J., Rao, M. \& {Aditi Simha}, R.} 2013 Hydrodynamics of soft active matter.
  {\em Rev. Mod. Phys.\/} {\bf 85}, 1143--1189.

\bibitem[Marenduzzo {\em et~al.\/}(2007{\natexlab{{\em a\/}}})Marenduzzo,
  Orlandini, Cates \& Yeomans]{Marenduzzo07b}
{\sc Marenduzzo, D., Orlandini, E., Cates, M. \& Yeomans, J.}
  2007{\natexlab{{\em a\/}}} Steady-state hydrodynamic instabilities of active
  liquid crystals: Hybrid lattice {Boltzmann} simulations. {\em Phys. Rev. E\/}
  {\bf 76}, 031921.

\bibitem[Marenduzzo {\em et~al.\/}(2007{\natexlab{{\em b\/}}})Marenduzzo,
  Orlandini \& Yeomans]{Marenduzzo07a}
{\sc Marenduzzo, D., Orlandini, E. \& Yeomans, J.} 2007{\natexlab{{\em b\/}}}
  Hydrodynamics and rheology of active liquid crystals: A numerical
  investigation. {\em Phys. Rev. Lett.\/} {\bf 98}, 118102.

\bibitem[Nash {\em et~al.\/}(2010)Nash, Adhikari, Tailleur \& Cates]{Nash10}
{\sc Nash, R.~W., Adhikari, R., Tailleur, J. \& Cates, M.~E.} 2010
  Run-and-tumble particles with hydrodynamics: Sedimentation, trapping, and
  upstream swimming. {\em Phys. Rev. Lett.\/} {\bf 104}, 258101.

\bibitem[Nitsche \& Brenner(1990)]{Nitsche90}
{\sc Nitsche, J.~M. \& Brenner, H.} 1990 On the formulation of boundary
  conditions for rigid non spherical {Brownian} particles near solid walls:
  Applications to orientation-specific reactions with immobilized enzymes. {\em
  J. Colloid Interface Sci.\/} {\bf 138}, 21--41.

\bibitem[Ravnik \& Yeomans(2013)]{Ravnik13}
{\sc Ravnik, M. \& Yeomans, J.~M.} 2013 Confined active nematic flow in
  cylindrical capillaries. {\em Phys. Rev. Lett.\/} {\bf 110}, 026001.

\bibitem[Riedel {\em et~al.\/}(2005)Riedel, Kruse \& Howard]{Riedel05}
{\sc Riedel, I.~H., Kruse, K. \& Howard, J.} 2005 A self-organized vortex array
  of hydrodynamically entrained sperm cells. {\em Science\/} {\bf 309},
  300--303.

\bibitem[Rothschild(1963)]{Rothschild63}
{\sc Rothschild, L.} 1963 Non-random distribution of bull spermatozoa in a drop
  of sperm suspension. {\em Nature\/} {\bf 198}, 1221--1222.

\bibitem[Rusconi {\em et~al.\/}(2014)Rusconi, Guasto \& Stocker]{Rusconi14}
{\sc Rusconi, R., Guasto, J.~S. \& Stocker, R.} 2014 Bacterial transport
  suppressed by fluid shear. {\em Nature Phys.\/} {\bf 10}, 212--217.

\bibitem[Rusconi {\em et~al.\/}(2010)Rusconi, Lecuyer, Guglielmini \&
  Stone]{Rusconi10}
{\sc Rusconi, R., Lecuyer, S., Guglielmini, L. \& Stone, H.~A.} 2010 Laminar
  flow around corners triggers the formation of biofilm streamers. {\em J. R.
  Soc. Interface\/} {\bf 7}, 1293--1299.

\bibitem[Saintillan \& Shelley(2008{\natexlab{{\em a\/}}})]{SS2008a}
{\sc Saintillan, D. \& Shelley, M.~J.} 2008{\natexlab{{\em a\/}}} Instabilities
  and pattern formation in active particle suspensions: Kinetic theory and
  continuum particle simulations. {\em Phys. Rev. Lett.\/} {\bf 100}, 178103.

\bibitem[Saintillan \& Shelley(2008{\natexlab{{\em b\/}}})]{SS2008b}
{\sc Saintillan, D. \& Shelley, M.~J.} 2008{\natexlab{{\em b\/}}}
  Instabilities, pattern formation, and mixing in active suspensions. {\em
  Phys. Fluids\/} {\bf 20}, 123304.

\bibitem[Saintillan \& Shelley(2013)]{SS2013}
{\sc Saintillan, D. \& Shelley, M.~J.} 2013 Active suspensions and their
  nonlinear models. {\em C. R. Physique\/} {\bf 14}, 497--517.

\bibitem[Schiek \& Shaqfeh(1995)]{Schiek95}
{\sc Schiek, R.~L. \& Shaqfeh, E. S.~G.} 1995 A nonlocal theory for stress in
  bound, {Brownian} suspensions of slender, rigid fibres. {\em J. Fluid
  Mech.\/} {\bf 296}, 271--324.

\bibitem[Sokolov {\em et~al.\/}(2010)Sokolov, Apodaca, Grzybowski \&
  Aranson]{Sokolov10}
{\sc Sokolov, A., Apodaca, M.~M., Grzybowski, B.~A. \& Aranson, I.~S.} 2010
  Swimming bacteria power microscopic gears. {\em Proc. Natl. Acad. Sci. USA\/}
  {\bf 107}, 969--974.

\bibitem[Spagnolie \& Lauga(2012)]{Spagnolie12}
{\sc Spagnolie, S.~E. \& Lauga, E.} 2012 Hydrodynamics of self-propulsion near
  boundaries: predictions and accuracy of far-field approximations. {\em J.
  Fluid Mech.\/} {\bf 700}, 105--147.

\bibitem[Suarez \& Pacey(2006)]{Suarez06}
{\sc Suarez, S.~S. \& Pacey, A.~A.} 2006 Sperm navigation in the female
  reproductive tract. {\em Human Reproduction Update\/} {\bf 12}, 23--37.

\bibitem[Subramanian \& Koch(2009)]{Subra09}
{\sc Subramanian, G. \& Koch, D.~L.} 2009 Critical bacterial concentration for
  the onset of collective swimming. {\em J. Fluid Mech.\/} {\bf 632}, 359--400.

\bibitem[Takagi {\em et~al.\/}(2013)Takagi, Braunschweig, Zhang \&
  Shelley]{Takagi13}
{\sc Takagi, D., Braunschweig, A., Zhang, J. \& Shelley, M.~J.} 2013 Dispersion
  of self-propelled rods undergoing fluctuation-driven flips. {\em Phys. Rev.
  Lett.\/} {\bf 110}, 038301.

\bibitem[Takagi {\em et~al.\/}(2014)Takagi, Palacci, Braunschweig, Shelley \&
  Zhang]{Takagi14}
{\sc Takagi, D., Palacci, J., Braunschweig, A., Shelley, M. \& Zhang, J.} 2014
  Hydrodynamic capture of microswimmers into sphere-bound orbits. {\em Soft
  Matter\/} {\bf 10}, 1784--1789.

\bibitem[Voituriez {\em et~al.\/}(2005)Voituriez, Joanny \& Prost]{Voituriez05}
{\sc Voituriez, R., Joanny, J.~F. \& Prost, J.} 2005 Spontaneous flow
  transition in active polar gels. {\em Europhys. Lett.\/} {\bf 70}, 404--410.

\bibitem[Wioland {\em et~al.\/}(2013)Wioland, Woodhouse, Dunkel, Kessler \&
  Goldstein]{Wioland13}
{\sc Wioland, H., Woodhouse, F.~G., Dunkel, J., Kessler, J.~O. \& Goldstein,
  R.~E.} 2013 Confinement stabilizes a bacterial suspension into a spiral
  vortex. {\em Phys. Rev. Lett.\/} {\bf 110}, 268102.

\bibitem[Woolley(2003)]{Woolley03}
{\sc Woolley, D.~M.} 2003 Motility of spermatozoa at surfaces. {\em
  Reproduction\/} {\bf 126}, 259--270.

\bibitem[{Z\"ottl} \& Stark(2012)]{Zottl12}
{\sc {Z\"ottl}, A. \& Stark, H.} 2012 Nonlinear dynamics of a microswimmer in
  {Poiseuille} flow. {\em Phys. Rev. Lett.\/} {\bf 108}, 218104.

\bibitem[{Z\"ottl} \& Stark(2013)]{Zottl13}
{\sc {Z\"ottl}, A. \& Stark, H.} 2013 Periodic and quasiperiodic motion of an
  elongated microswimmer in {Poiseuille} flow. {\em Eur. Phys. J. E\/} {\bf
  36}, 4.

\end{thebibliography}

\end{document}